\documentclass[preprint2,times,twocolappendix]{aastex6}

\usepackage{amsmath,amssymb,natbib}
\usepackage{microtype}
\usepackage{hyperref}
\usepackage{graphicx}

\newcommand{\project}[1]{\textsl{#1}}
\newcommand{\code}[1]{\textit{#1}}
\newcommand{\hscPipe}{\code{hscPipe}}
\newcommand{\sextractor}{\code{SExtractor}}
\newcommand{\area}{200}
\newcommand{\ngal}{781}
\newcommand{\sbunit}{mag~arcsec$^{-2}$}
\newcommand{\HI}{H{\sc\,i}}
\newcommand{\mueff}{$\bar{\mu}_\mathrm{eff}(g)$}
\newcommand{\survey}{HSC-SSP}

\newcommand{\jpg}[1]{#1}
\newcommand{\resp}[1]{#1}

\definecolor{citecolor}{rgb}{0,0.1,0.43}
\definecolor{linkcolor}{HTML}{DC322F}
\hypersetup{backref,breaklinks,colorlinks,
  urlcolor=citecolor,linkcolor=linkcolor,citecolor=citecolor}
\graphicspath{{figures/}}

\shorttitle{HSC-SSP Low-Surface-Brightness Galaxies}
\shortauthors{Greco et al.}

\begin{document}\sloppy\sloppypar\raggedbottom\frenchspacing 

\title{Illuminating Low-Surface-Brightness Galaxies with the Hyper Suprime-Cam Survey}
\author{Johnny~P.~Greco\altaffilmark{\pu}, 
        Jenny~E.~Greene\altaffilmark{\pu},
        Michael~A.~Strauss\altaffilmark{\pu},
        Lauren~A.~MacArthur\altaffilmark{\pu},
        Xzavier Flowers\altaffilmark{\pu,\fit},
        Andy~D.~Goulding\altaffilmark{\pu},
        Song Huang\altaffilmark{\santa},
        Ji Hoon Kim\altaffilmark{\subaru},
        Yutaka Komiyama\altaffilmark{\naoj,\sokendai},
        Alexie Leauthaud\altaffilmark{\santa},
        Lukas Leisman\altaffilmark{\VU},
        Robert~H.~Lupton\altaffilmark{\pu},
        Crist{\'o}bal Sif{\'o}n\altaffilmark{\pu},
        Shiang-Yu Wang\altaffilmark{\iaa}
}
\email{jgreco@astro.princeton.edu}

\newcommand{\pu}{1}
\newcommand{\santa}{2}
\newcommand{\naoj}{3}
\newcommand{\sokendai}{4}
\newcommand{\subaru}{5}
\newcommand{\iaa}{6}
\newcommand{\VU}{7}
\newcommand{\fit}{8}
\altaffiltext{\pu}{Department of Astrophysical Sciences, 
Princeton University, Princeton, NJ 08544, USA}
\altaffiltext{\santa}{Department of Astronomy and Astrophysics, 
University of California, Santa Cruz, 
1156 High Street, Santa Cruz, CA 95064 USA}
\altaffiltext{\naoj}{National Astronomical Observatory of Japan, 
2-21-1 Osawa, Mitaka, Tokyo 181-8588, Japan}
\altaffiltext{\sokendai}{Department of Astronomy, School of Science, 
Graduate University for Advanced Studies (SOKENDAI), 2-21-1, Osawa, Mitaka, 
Tokyo 181-8588, Japan}
\altaffiltext{\subaru}{Subaru Telescope, National Astronomical 
Observatory of Japan, 650 N Aohoku Pl, Hilo, HI 96720}
\altaffiltext{\iaa}{Institute of Astronomy and Astrophysics, 
Academia Sinica, P. O. Box 23-141, Taipei 106, Taiwan}
\altaffiltext{\VU}{Department of Physics and Astronomy, 
Valparaiso University, Valparaiso, IN 46383, USA}
\altaffiltext{\fit}{Florida Institute of Technology
150 W. University Blvd
Melbourne, Florida 32901}

\begin{abstract}
We present a catalog of extended low-surface-brightness galaxies (LSBGs)
identified in the Wide layer of the Hyper Suprime-Cam Subaru Strategic Program
(HSC-SSP). Using the first ${\sim}$200~deg$^2$ of the survey, 
we have uncovered 781 LSBGs, spanning red ($g-i\geq0.64$) and blue
($g-i<0.64$) colors and a wide range of morphologies. Since we focus on  
extended galaxies ($r_\mathrm{eff}=2.5$-$14\arcsec$), our sample is likely 
dominated by low-redshift objects. We define LSBGs to have mean surface brightnesses
$\bar{\mu}_\mathrm{eff}(g)>24.3$~mag~arcsec$^{-2}$, which allows nucleated
galaxies into our sample. As a result, the central surface brightness
distribution spans a wide range of $\mu_0(g)=18$-$27.4$~mag~arcsec$^{-2}$, with
50\% and 95\% of galaxies fainter than 24.3 and 22~mag~arcsec$^{-2}$,
respectively. Furthermore, the surface brightness distribution is a
strong function of color, with the red distribution being much broader and
generally fainter than that of the blue LSBGs, and this trend shows a clear
correlation with galaxy morphology. Red LSBGs typically have smooth light
profiles that are well-characterized by single-component S\'{e}rsic functions.
In contrast, blue LSBGs tend to have irregular morphologies and show evidence
for ongoing star formation. We crossmatch our sample with existing optical,
H{\sc\,i}, and ultraviolet catalogs to gain insight into the physical nature of
the LSBGs. We find that our sample is diverse, ranging from dwarf
spheroidals and ultra-diffuse galaxies in nearby groups to gas-rich 
irregulars to giant LSB spirals, demonstrating the potential of the 
HSC-SSP to provide a truly unprecedented view of the LSBG population.  
\end{abstract}


\section{Introduction}

Low-surface-brightness galaxies (LSBGs) are a significant component of the
galaxy population \citep{McGaugh:1995ab,Dalcanton:1997aa}, \resp{which spans
a broad range of galaxy properties} and environments 
\citep[e.g.,][]{Bothun:1987aa,Impey:1996aa, ONeil:1997aa, Zucker:2006aa,
  McConnachie:2012aa,McGaugh:1995aa,Beijersbergen:1999aa,van-Dokkum:2015aa}.
Furthermore, many of the most pressing problems currently facing the $\Lambda$
Cold Dark Matter ($\Lambda$CDM) paradigm were discovered (and may be resolved)
through studies of galaxies in the low-luminosity and/or LSB regime
\citep[e.g.,][]{Kauffmann:1993aa, Klypin:1999aa, Moore:1999aa,
  Boylan-Kolchin:2011aa}. Despite the importance of LSBGs (for a review
  of classical LSBGs, see \citealt{Impey:1997aa} and \citealt{Bothun:1997aa}),
their defining characteristic---central surface brightnesses that are fainter
than the night sky---makes them difficult to detect and study, which has led to
their underrepresentation in previous optical surveys, thus biasing our view
of the full galaxy population \citep{Disney:1976aa}. 

Indeed, uncovering the distribution of galaxies at ever-lower surface
brightnesses remains an active area of research \citep[e.g.,][]{Impey:1988aa,
  McGaugh:1996aa, Blanton:2005aa, Munoz:2015aa, van-der-Burg:2016aa,
  Roman:2017aa}. Modern wide-field surveys such as the Sloan Digital Sky Survey
\citep[SDSS;][]{York:2000aa} have enabled statistical studies of large samples
of LSBGs down to central surface brightnesses of $\mu_0(B)\sim24$~\sbunit\
\citep[e.g.,][]{Zhong:2008aa,Rosenbaum:2009aa,Galaz:2011aa}, and fainter limits
are continuously being reached through advances in observing strategies and
data reduction \citep[e.g.,][]{Blanton:2011aa, Ferrarese:2012aa, Duc:2015aa,
  Fliri:2016aa, Trujillo:2016aa}. Pushing to still lower surface brightnesses,
optical and \HI\ surveys have been combined to detect and characterize
populations of gas-rich LSBGs
\citep[e.g.,][]{Du:2015aa,Leisman:2017aa}. 

Small robotic telescopes optimized for LSB science
\citep[e.g.,][]{Martinez-Delgado:2010aa, Abraham:2014aa, Javanmardi:2016aa}
have recently joined the search for LSBGs, which notably resulted in the
discovery of a significant population of red, ultra-LSB galaxies within the
Coma cluster \citep{van-Dokkum:2015aa,Koda:2015aa,Yagi:2016aa}. With central
surface brightnesses $\mu_0(g)\gtrsim24$ \sbunit\ and effective radii
$r_\mathrm{eff}\gtrsim1.5$~kpc, these so-called ultra-diffuse galaxies (UDGs)
are remarkable in that they have stellar masses similar to dwarf galaxies
spread over diameters comparable to that of the Milky Way. While UDG-like
objects have been known to exist for decades \citep[e.g.,][]{Sandage:1984aa},
their unexpected abundance in the Coma cluster has reignited the search for
LSBGs, leading to the discovery of UDGs in environments ranging from dense
galaxy clusters to the field
\citep{Mihos:2015aa,Martinez-Delgado:2016aa,Merritt:2016aa,
  Roman:2017ab,van-der-Burg:2017aa,Bellazzini:2017aa,Leisman:2017aa}. 

In the $\Lambda$CDM framework, LSBGs naturally arise within dark matter
halos with high angular momentum \citep{Dalcanton:1997ab}. Therefore,
measurements of their number densities and kinematic properties can provide
powerful tests of cosmological models \citep[e.g.,][]{Ferrero:2012aa,
  Papastergis:2015aa}. For UDGs specifically, their number density as a
function of environment can be explained if they preferentially form in
dwarf-mass halos \jpg{($M_\mathrm{virial}\sim10^{10}~M_\odot$)} with higher
than average \jpg{angular momentum} \citep{Amorisco:2016aa}, a view which is
consistent with recent weak-lensing \citep{Sifon:2018aa} and \HI\
\citep{Leisman:2017aa} observations. Alternatively, \citet{van-Dokkum:2015aa}
suggested UDGs may be ``failed'' galaxies, with external mechanisms such as gas
stripping and/or extreme feedback processes suppressing the growth of normal
stellar populations at a given halo mass. This scenario also appears to be viable 
for at least some fraction of UDGs \citep[e.g.,][]{van-Dokkum:2016aa}. Of
course, it is crucial to understand how UDGs fit into the broader context of
the LSBG population and whether these galaxies have different formation paths
as a function of physical properties (e.g., size, mass, and color) and
environment.  

The new generation of wide-field optical imaging surveys such as the Dark Energy Survey
\citep{DES:2016aa}, the Kilo-Degree Survey \citep{de-Jong:2015aa}, our ongoing
Hyper Suprime-Cam Subaru Strategic Program \citep[\survey;][]{Aihara:2018aa},
and ultimately the Large Synoptic Survey Telescope
\citep[LSST;][]{Ivezic:2008aa} will extend our census of the galaxy population
to lower surface brightnesses than has previously been possible over large areas of
the sky. Although it will require dedicated LSB-optimized reduction and
analysis efforts, these surveys will produce statistical samples of LSBGs that
span all halo environments, opening a new window into their formation and
evolution and providing ideal systems to test the predictions of $\Lambda$CDM. 

In this work, we present initial results from our search for LSBGs with the
\survey. We focus on angularly extended (and therefore primarily low-$z$)
LSBGs. We define and select LSBGs to have $g$-band mean surface brightnesses 
within their circularized effective radii \mueff$>24.3$~\sbunit, \resp{which is 
equivalent to $\mu_0(g)>23.5$~\sbunit\ for a S\'{e}rsic surface-brightness
profile \citep{Sersic:1968aa} with S\'{e}rsic index $n=0.8$.}

The paper is organized as follows. In Section~\ref{sec:survey}, we discuss the
Wide layer of the \survey\ on which our study is based. We then describe our
LSBG source-detection pipeline in Section~\ref{sec:pipeline}. In
Section~\ref{sec:sample}, we present our galaxy sample. In
Section~\ref{sec:xmatch}, we crossmatch our catalog with previous work to gain
insight into the span of physical properties within our sample. We conclude
with a summary and outlook in Section~\ref{sec:summary}. In addition, we give
examples of non-galactic LSB sources detected by our pipeline in the Appendix. 

For all relevant calculations, we assume a standard cosmology with $H_0=70$
km~Mpc$^{-1}$, $\Omega_m=0.3$, and $\Omega_\Lambda=0.7$. All magnitudes
presented in this paper use the AB system \citep{Oke:1983aa}. Unless stated
otherwise, we correct for Galactic extinction using the $E(B-V)$ values from the
dust map of \citet{Schlegel:1998aa} and the recalibration from
\citet{Schlafly:2011aa}.

\section{The Wide Layer of \survey}\label{sec:survey}

Our search is based on the first ${\sim}\area$~deg$^2$ from the Wide layer of
the \survey\ (internal data release S16A), an ambitious 300-night imaging
survey using the Hyper Suprime-Cam \citep{Miyazaki:2018aa} on the 8.2-meter
Subaru Telescope. Upon completion, the Wide layer will cover
${\sim}1400$~deg$^2$ in five broad bands ($grizy$), achieving a depth of
$i\sim26$~mag ($5\sigma$ point-source detection) with a median seeing of
$0\farcs6$. An overview of the full survey design is given in
\citet{Aihara:2018aa}, and the first public data release, which covers
${\sim}100$~deg$^2$, is described in \citet{Aihara:2018ab}. Here, we will
briefly summarize some key aspects of the survey that are relevant to our
search. 

The \survey\ data are processed using an open-source software package,
\hscPipe, which builds upon the pipeline being built for the LSST Data
Management system\footnote{\url{http://dm.lsst.org}}
\citep{Axelrod:2010aa, Juric:2015aa}. The software pipeline is described in
detail by \citet{Bosch:2018aa}.  For our search, we work with the fully
reduced, sky-subtracted coadd images.  \hscPipe\ divides these images into
equi-area rectangular regions called \project{tracts}, which are predefined as
iso-latitude tessellations. Each tract covers $1.7$~deg$^2$ of sky, and
neighboring tracts have an overlap of ${\sim}1^\prime$ near the equatorial
fields. The tracts are further divided into $9\times9$ grids of
\project{patches}. Each patch is composed of $4200\times4200$ pixels
(${\sim}12^\prime$ on a side), with 100-pixel (${\sim}17\arcsec$) overlaps
between adjacent patches. 

\resp{Since we are searching for galaxies that are by definition fainter
than the night sky, the quality of the sky subtraction is important for 
our search. As described in \citet{Bosch:2018aa}, \code{hscPipe} estimates 
the sky on a CCD-by-CCD basis by averaging pixels in $128\times128$ pixel grids, 
where pixels belonging to detected objects are ignored. The average pixel values
are then fit with a 6th-order two-dimensional Chebyshev polynomial, which 
is used to subtract a smooth, slowly-varying flux distribution from the image.
This algorithm is known to oversubtract the background around large extended 
objects on the sky ($>1\arcmin$) such as bright, nearby galaxies 
\citep{Aihara:2018ab}. Therefore, our LSBG search will likely be biased against 
detecting galaxies projected within a few effective radii of such objects.  
Implementing a custom sky-subtraction algorithm across the entire \survey\ 
footprint is beyond the scope of this paper, but we note that a new and improved
sky-subtraction algorithm that uses the entire HSC field of view has been 
developed and will be implemented in future \survey\ data releases.}

We carry out our search on a patch-by-patch basis; in
total, we search 11,176 patches, which after accounting for overlaps and
masking, reduces to ${\sim}\area$~deg$^2$. We restrict our search to
patches that have been observed to the full Wide layer depth in $g$, $r$, and
$i$ as of May 2016.  We note that we only work with $gri$, since requiring all
five bands ($grizy$) at the full Wide layer depth would limit our survey area.
Studies using the full spectral range are forthcoming.

\begin{figure*}[ht!]
\centering
\includegraphics[width=\textwidth]{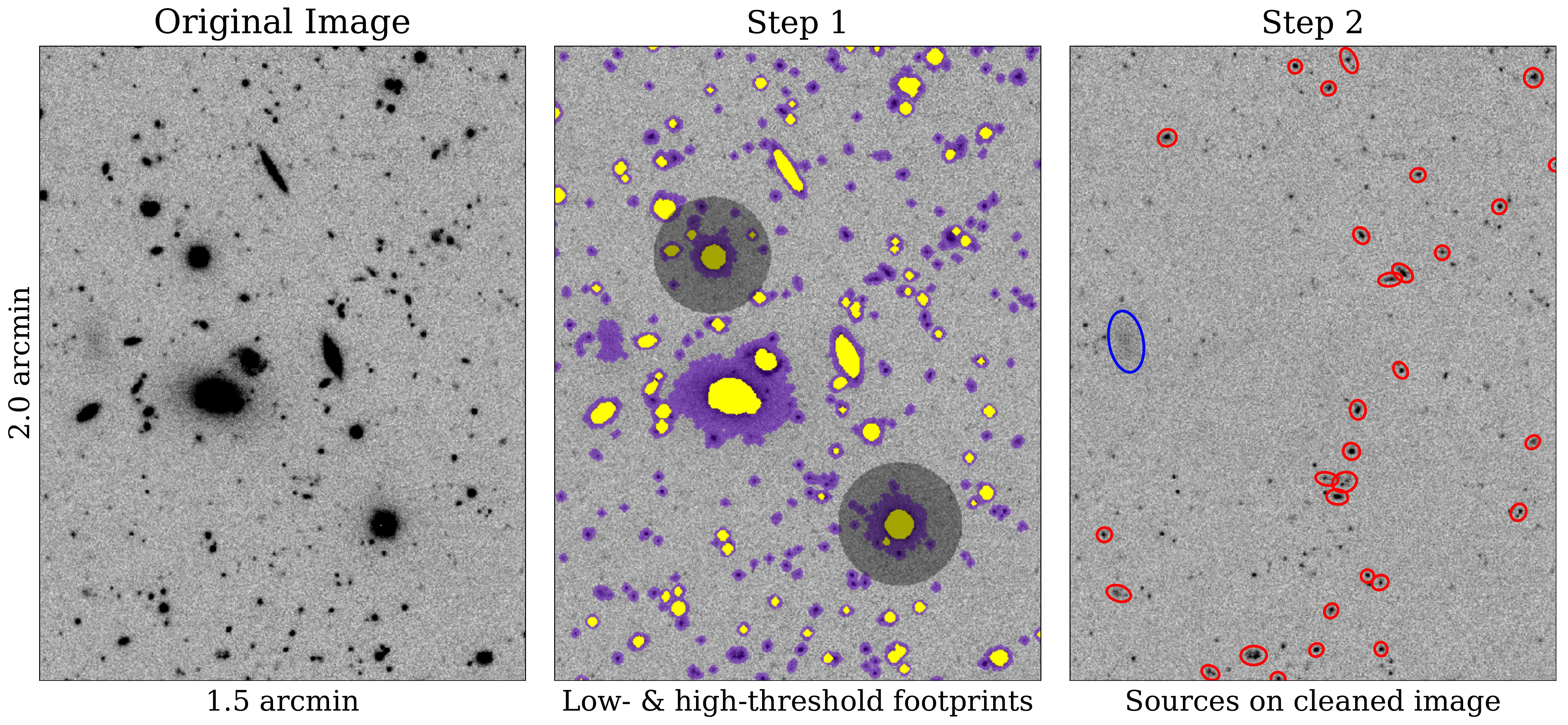}
\caption{
  \survey\ $i$-band images demonstrating the first two steps of our pipeline.
  We show a small section of a single patch for clarity. The original image is
  shown in the left panel, with the dimensions of the cutouts indicated on the
  axes. In the middle panel, we demonstrate how we find and remove bright
  sources and their associated diffuse light using multiple levels of
  thresholding.  We show overlays of the high-threshold footprints (yellow),
  which identify bright sources, and the low-threshold footprints (purple),
  which identify the associated diffuse light. The two shaded circles are from
  the bright-object mask supplied by \hscPipe\
  \citep{Coupon:2018aa,Mandelbaum:2018aa}. We replace a low-threshold footprint
  with sky noise if more than 15\% of its pixels are above the high threshold
  value. We also replace the \hscPipe\ bright-object mask with sky noise, producing
  a final ``cleaned'' image.  In the right panel, we show ellipses on 
  sources identified within the cleaned image. Note the large LSB source with
  the blue ellipse, which is the only detected object that passes our size
  cut (see Section~\ref{sec:selection}). The faint sources apparent in the image
  that were not detected (i.e., those without an ellipse) had too few connected 
  pixels above the required threshold.
}
\label{fig:pipeline} 
\end{figure*} 

\section{Source-Detection Pipeline}\label{sec:pipeline}

Currently, \hscPipe\ is optimized to identify faint, small (several
  arcseconds radius) objects, which are characteristic of galaxies in the
  distant universe---identifying these objects is one of the main drivers of
  the \survey.  Unfortunately, this is not optimal for the detection of
  extended LSBGs; \hscPipe\ tends to decompose these single extended systems
  into a set of child objects, so-called ``shredding.'' This problem also
  affects the measurements of bright, nearby galaxies, which often have
  significant sub-structure that leads to over-deblending
  \citep{Aihara:2018ab}. Galaxy shredding is well known to exist in 
  the Sloan Digital Sky Survey \citep[SDSS; e.g.,][]{Kniazev:2004aa, Fukugita:2007aa}, 
  but it is much more pronounced at the depths of the \survey. 

A typical extended LSBG in the current \survey\ catalog will be shredded into
many ($\gtrsim10$) child objects and a parent object centered on the brightest
peak within the source footprint, which will not necessarily be associated with
the luminosity- or position-weighted center of the galaxy. Furthermore, the
faint integrated light from LSBGs is often dominated by background/foreground
objects that may be assigned as the galaxy's center if they fall within the
extended LSB footprint.  As a result, shape and surface-brightness measurements
of LSBGs in the \survey\ catalog are often spurious, making it difficult to
construct a robust selection of such objects. To address this issue, we have
developed a custom pipeline to perform our search. 
 
Our source-extraction
pipeline is open-source\footnote{\url{https://github.com/johnnygreco/hugs}} and
is currently under active development. Our software is based primarily on the
LSST codebase, which provides a powerful suite of tools for
optical/near-infrared source detection and photometry. We additionally use
\sextractor\footnote{\url{https://www.astromatic.net/software/sextractor}} 
\citep{Bertin:1996aa} in our pipeline's final detection step to
estimate initial galaxy parameters for selection and
\code{imfit}\footnote{\url{http://www.mpe.mpg.de/~erwin/code/imfit/}} 
\citep{Erwin:2015aa} to refine our parameter estimates.
 
In this section, we describe the major steps of our pipeline.  As noted in
Section~\ref{sec:survey}, our search is carried out using the coadd images from
the Wide layer of \survey, which have been fully reduced (including sky
subtraction) by \hscPipe. We perform our detection in the $i$-band (typical
seeing ${\sim}0\farcs6$), and require that all sources also be detected in the
$g$-band (typical seeing ${\sim}0\farcs7$) to reduce the number of false
positives associated with optical artifacts such as ghosts or scattered light
from bright stars. The image-processing steps described below are applied to
$i$-band images unless stated otherwise. 

For the reader only interested in the broad outline of our pipeline, the 
major steps (which we describe in detail below) are the following:
\begin{enumerate}
  \item {\bf Remove bright sources and associated diffuse light.} LSB
    light associated with nearby, bright sources can mimic the signal of our
    objects of interest. We detect these sources using multiple levels
    of thresholding and replace their footprints with sky noise. 
  \item {\bf Source extraction.} We smooth the ``cleaned'' images produced by
    the previous step with a fairly large Gaussian kernel (FWHM=1\arcsec) and
    extract sources using \sextractor\ that are composed of 100 or more connected 
    pixels (1 pixel = $0\farcs168$) with a low detection threshold of
    0.7$\sigma$ per pixel. This produces a sample of ${\sim}7\times10^6$ 
    unique sources, the vast majority of which are small and faint. 
  \item {\bf Initial sample selection.} We then make selection cuts on the
    parameters measured by \sextractor. By far the most powerful cut we apply
    is $r_{1/2}>2\farcs5$, where $r_{1/2}$ is the half-light radius. This cut
    alone reduces our sample to 20,838 sources. We also apply reasonable color
    cuts based on \sextractor\ aperture photometry. 
  \item {\bf Galaxy Modeling.} We model the two-dimensional surface brightness
    profiles of the remaining candidates using {\it imfit}. We then make a second
    selection based on the mean ($g$-band) surface brightness within the
    circularized effective radius \mueff\ as measured by {\it imfit}. In
    addition, we remove likely astrophysical false positives by comparing the
    measurements of \sextractor\ and \code{imfit}. This step reduces our sample
    to 1521 candidate LSBGs. 
  \item  {\bf Visual inspection.} We visually inspect all the remaining
    candidates and remove any obvious false positives, which are dominated by
    the blending of point-like sources with background/foreground diffuse light
    (e.g., from a nearby bright star, massive low-$z$ galaxy, or Galactic cirrus),
    which were not removed by Step 1. After this step, we are left with a final
    sample of \ngal\ LSBGS. 
 \end{enumerate}

\subsection{Step 1: Bright Sources and Associated Diffuse Light}

The first step of our pipeline is to remove bright sources and their associated
diffuse light (e.g., the LSB outskirts of giant elliptical galaxies). While the
former may in principle be removed at the catalog level, the latter tends to be
shredded into individual objects whose measured properties are very similar to
our objects of interest. This problem is often addressed by using multiple
\sextractor\ runs with different configurations that are optimized to detect
high-surface-brightness (HSB) and LSB sources separately
\citep[e.g.,][]{Rix:2004aa, Barden:2012aa, Prescott:2012aa}. The resulting
catalogs can then be crossmatched to build a final, more complete catalog. One
benefit of using the LSST codebase is that we have complete
control of each image-processing step. This allows us to perform multiple
levels of thresholding without having to build a full catalog at each level, as
is necessary when using multiple \sextractor\ runs. 

For each patch (${\sim}12^\prime$ on a side; see Section~\ref{sec:survey}), we
use the following steps to remove bright sources and their associated diffuse
light. We start by smoothing the image with a circular Gaussian matched to the
rms width of the point-spread function (PSF). Image convolution maximizes the
ratio of a source's peak signal to the local noise level
\citep[e.g.,][]{Irwin:1985aa, Akhlaghi:2015aa}, and the PSF scale is formally
optimal for the idealized case of detecting isolated point sources
\citep{Bosch:2018aa}. Additionally, this is the smallest scale at which we can
detect real astrophysical sources, so it reduces the impact of pixels with high
noise fluctuations. Next, we find very bright sources by flagging all pixels
that are at least $28\sigma$ above the global background level for each patch;
for a typical patch, this corresponds to the brightest ${\sim}2\%$ of all
pixels. The background and its variance are estimated using several iterations
of sigma clipping. We then identify LSB structures as collections of 20 or more
pixels that are at least $3\sigma$ above the background level; \resp{this 
LSB image thresholding is carried out on the identical image as the previous HSB 
thresholding.} Finally, we associate LSB structures with bright 
sources if more than 15\% of their pixels are above the high threshold used 
during the first round of thresholding.  

The numbers presented above (28$\sigma$, 3$\sigma$, and 15\%) were chosen after
many iterations of trial and error. There is a trade-off between aggressively
removing all LSB pixels that are physically associated with HSB sources and
inadvertently removing LSBGs that happen to be near (in projection or
physically) HSB sources. The middle panel of Figure~\ref{fig:pipeline} shows an
example image with low- (purple) and high-threshold (yellow) footprints
overlaid. 

We generate a final bright object mask by taking the union of the source
footprints (bright sources + associated diffuse light) and the bright object
masks provided by \hscPipe---we use an early version of the masks called
\textit{Sirius} \citep{Mandelbaum:2018aa}, which has since been superseded by
\textit{Arcturus} \citep{Coupon:2018aa}. The two shaded circles in the middle
panel of Figure~\ref{fig:pipeline} show bright objects that were masked by
\hscPipe. We then replace masked pixels with sky noise generated from a
Gaussian distribution with zero mean (the images have been background
subtracted by \hscPipe) and standard deviation equal to the rms pixel value. We
use the same mask to perform noise replacement of pixels in the $g$-, $r$-, and
$i$-band images. An example ``cleaned'' $i$-band image is shown in the right
panel of Figure~\ref{fig:pipeline}.

\subsection{Step 2: Source Extraction}

The previous step of our pipeline produces ``cleaned'' images for each patch,
where bright sources and their associated diffuse light have been replaced by
sky noise.  Next, we perform source extraction and estimate an initial set of
galaxy parameters.  While it would be possible to use the LSST codebase for this
step, we found that the currently implemented shape-measurement algorithms tend
to fail for very LSB, extended targets. We therefore use \sextractor\ for this
task, which has been well tested in the LSB regime
\citep[e.g.,][]{Yagi:2016aa,van-der-Burg:2016aa}.

We perform our detections in the $i$-band but require they also be detected in
the $g$-band to reduce contamination from optical artifacts. Before detection,
we smooth the images with a Gaussian kernel with FWHM~$=1\arcsec$ (nearly twice 
the typical $i$-band seeing) to boost our LSB 
sensitivity. We experimented with larger kernel sizes up to FWHM~$=6\arcsec$ but 
found that the
increase in contamination from blends was not worth the gain in sensitivity.
Blends of faint galaxies and/or distant galaxy groups are a major source of
contamination for any deep diffuse-galaxy search \citep[e.g.,][]{Koda:2015aa,
  Sifon:2018aa}. The faintness of these sources means they will not be masked
by the previous step of our pipeline, and if detected as single objects, their
sizes will be biased high, and their central surface brightnesses will be
biased low, making them look like the objects we are searching for. 

Although we are working on background-subtracted images, we re-estimate the
{\it local} background using a mesh size of 128 pixels $\approx22\arcsec$. 
We use a low detection threshold of 0.7$\sigma$ per pixel above the local 
background level; \resp{at lower thresholds, contamination 
(e.g., from the LSB outskirts of bright galaxies or diffuse light 
from stars) becomes too overwhelming.}
We require all detections be composed of at
least 100 contiguous pixels.  We require such large footprints because of 
our low detection threshold and
because we are primarily interested in finding very extended LSBGs. \resp{
For HSC-SSP images, an area of 100 pixels corresponds to a circular region of 
radius ${\sim}$1-2 times the FWHM of the PSF.}

We perform aperture photometry (to be used for selection) in the $i$-band using
an aperture size of 2.5 times (the default value for \sextractor\ \textsc{auto}
parameters) the Kron radius \citep{Kron:1980aa}. To measure colors (again, only
for selection purposes), we apply the $i$-band apertures to the $g$- and
$r$-band images---the photometry is forced. 

For a typical patch (${\sim}12^\prime\times12^\prime$ region of sky), the above
procedure generates a catalog of ${\sim}900$ objects, most of which can be
rejected due to their small angular size. In the rightmost panel of
Figure~\ref{fig:pipeline}, the blue ellipse shows a detection of an ultra-LSB
source. Note its large angular size compared to the other sources. In total,
this step generated a catalog of ${\sim}7\times10^6$ unique sources after 
removing duplicate detections due to overlaps between patches.

\begin{figure*}[ht!]
  \centering
  \includegraphics[width=\textwidth]{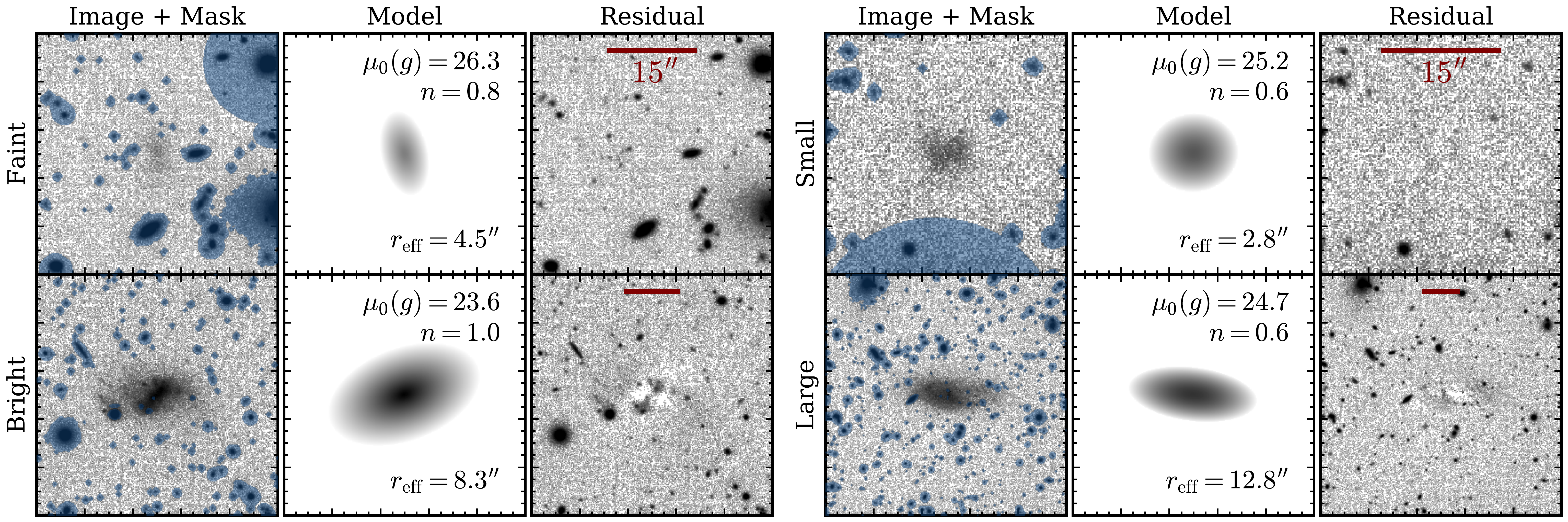}
  \caption{
    Example S\'{e}rsic function fitting results. We show $i$-band cutouts that
    are $12\times r_{1/2}$ on a side, where $r_{1/2}$ is the half-light radius
    of the associated galaxy, as measured by \sextractor\ (i.e., not
    $r_\mathrm{eff}$, which is measured by {\it imfit}). The images are shown
    with a logarithmic stretch. The left columns show the original galaxy
    images with overlays of the object masks, the middle columns show the
    best-fit models with the main model parameters indicated for each galaxy,
    and the right columns show the residual images. We show results for 4
    galaxies from our final sample, which are faint (top left), bright (bottom left), 
    small (top right), and large (bottom right) with respect to the full sample. The
    angular  scales are indicated at the top of each residual image.
  } 
  \label{fig:fits} 
\end{figure*}

\subsection{Step 3: Initial Sample Selection}\label{sec:selection}

We now seek a set of selection cuts that will reduce our sample to a manageable size
for more detailed galaxy modeling. To accomplish this, we use size and color
measurements from \sextractor. In particular, we keep sources that have
$i$-band half-light radii, as measured by \sextractor\ via the
\textsc{flux\_radius} parameter, within $2\farcs5 < r_{1/2} < 20\arcsec$. The
lower bound is due to our interest in extended galaxies; \jpg{for objects
within $z\sim0.01$-0.03 (approximate range of our closest objects 
with known redshifts; see Section~\ref{sec:xmatch}),
$2\farcs5$ corresponds to $r_{1/2}\sim0.5$-1.5~kpc.} \resp{Since we perform our 
search on a patch-by-patch basis, the upper bound is set by the 17\arcsec\ 
overlap between patches (see Section~\ref{sec:survey}). We inspected detections
larger than this scale, and they are rare and generally spurious. 
The minimum radius cut} is by far our most powerful selection
criterion, reducing the number of sources from $7\times10^6$ to 20,838.

We further require the \sextractor-measured colors to be within the color box
defined by 
\begin{align*}
-0.1  &< g-i < 1.4,\\
(g-r) &> 0.7\cdot(g-i) - 0.4,\ \mathrm{and}\\
(g-r) &< 0.7\cdot(g-i) + 0.4. 
\end{align*}
These color requirements were determined empirically using our full
catalog of objects and primarily remove spurious detections due to optical
artifacts detected in all bands and blends of high-redshift galaxies. This
color box is conservative with respect to what we expect from red and
blue LSBGs \citep[e.g.,][]{Roman:2017aa,Roman:2017ab, Geha:2017aa}, 
\resp{as well as with respect to reasonable assumptions about  
stellar populations (as we show in Section~\ref{sec:colors}).}
This cut reduces our sample to 14,069 sources.

\subsection{Step 4: Galaxy Modeling}\label{sec:modeling}
 
For the LSBG candidates that remain after the selection cuts of
Section~\ref{sec:selection}, we next refine our galaxy parameter measurements
using \code{imfit}. For each candidate, we extract a square cutout image that
is $12\times r_{1/2}$ on a side and centered on the centroid determined by
\sextractor.  The cutout images have been fully reduced by \hscPipe\ (including
background subtraction). We mask objects that are not associated with the
galaxy's smooth light profile using a combination of the \hscPipe\ bright
object masks and an automatically generated object mask using
\code{sep}\footnote{\url{https://sep.readthedocs.io}} 
\citep{Barbary:2016aa}, a source extraction Python package 
based on the core algorithms of \sextractor. We run our masking software 
on all bands individually and take the union to form a final mask. 
When performing the fits,
we use the variance images and PSF measurements provided by \hscPipe. We model
the surface-brightness distribution as a two-dimensional, PSF-convolved
S\'{e}rsic function. We first perform fits in the
$i$-band, which by design typically has the best quality data for \survey. To
measure colors, we fit the $g$- and $r$-band images with all parameters except
for the amplitude fixed at their $i$-band values; note this assumes no color
gradients. In Figure~\ref{fig:fits}, we show example fit results for LSBGs that
are faint, bright, small, and large with respect to the median of our full sample.  

Since the characteristic central surface brightness of our sample is much
fainter than the brightness of the night sky, the measured galaxy parameters
are sensitive to the assumed sky value, which is estimated by \hscPipe. We
quantify the impact of this uncertainty by performing the above fitting
procedure a second time with an additional free parameter for the sky value in
each cutout. We then compare the best-fit total magnitudes, central surface
brightnesses ($\mu_0$), S\'{e}rsic indices ($n$), effective radii
($r_\mathrm{eff}$), and ellipticities ($\epsilon$) with those of our previous
models where the sky is fixed. \jpg{ We take the rms of the distributions of
  the parameter differences to be the statistical uncertainties associated with
  the sky estimate. For a small fraction of objects ($<\!3\%$), varying the sky
  value leads to significant differences in the parameters (particularly the
  effective radii). These objects tend to be either very LSB and in close
  proximity to a bright star or low-$z$ galaxy, or multi-component systems that
  are less well-described by a single S\'{e}rsic model; we sigma clip the
  distributions at $5\sigma$ to remove these rare cases before calculating the
  rms values. We add these values in quadrature with the formal fitting
  uncertainties estimated by \code{imfit} to derive final uncertainties, which
  range from ${\sim}10$-30\% in effective radii and ${\sim}0.2$-0.4 mag in
  total magnitudes.}

With refined galaxy parameter estimates from {\it imfit}, we next make a second
set of selection cuts. We keep candidates that satisfy the following:
\begin{enumerate}
  \item Effective radius, measured along the major axis, $r_\mathrm{eff} >
    2\farcs5$, since our focus is 
    on extended LSBGs. 
  \item $g$-band mean effective surface brightness within the circularized 
    effective radius $24.3 <
    \bar{\mu}_\mathrm{eff}(g) < 28.8$. This is equivalent to requiring central
    surface brightness within $23.5 < \mu_0(g) < 28$ for a S\'{e}rsic profile
    with $n=0.8$. Low-mass LSBGs typically have $n\sim1$
    \citep[e.g.,][]{Geha:2006aa,Roman:2017aa}. Our choice of $n=0.8$ is more
    inclusive than an exponential profile. We cut on \mueff\ rather than
    $\mu_0(g)$ to allow nucleated galaxies into our sample.
  \item Ellipticity $\epsilon < 0.7$. This cut removes nearly edge-on, high
    redshift galaxies, whose surface brightnesses appear low due to
    cosmological dimming, as well as some galaxy blends and linear optical
    artifacts that tend to have high ellipticity. While in principle this biases 
    our sample to some degree, we find that most of our candidates are 
    far less elongated. 
  \item We remove likely astrophysical false positives (e.g., blends and
    residual light from a bright galaxy that was not completely masked) by
    comparing the parameters measured by \sextractor\ and \code{imfit}.
    Specifically, we require that the difference between the centroids be less than
    4\arcsec, and the difference between \sextractor's {\sc mag\_auto}
    parameter and the total magnitude measured by \code{imfit} be less than
    1~mag in all bands ($gri$).
\end{enumerate}
The above selection produced a catalog of 1521 LSBG candidates.

\subsection{Step 5: Visual Inspection}

We visually inspect all of the LSBG candidates from the previous 4 steps and
remove any remaining false positives; ${\sim}50\%$ of the objects are
eliminated by this step. The dominant false positives are due to blends of
point-like sources superposed on extended halos of diffuse light (e.g., from a
bright star, giant low-$z$ galaxy, or Galactic cirrus). We can reduce our
contamination to ${\sim}25\%$ by being more aggressive during the masking in
Step~1; however, this comes at the cost of reducing our sample size by a factor
of 2. We choose to accept the extra contamination in favor of better
completeness. 

The first two authors independently performed the visual inspection.  We note
that it is difficult to consistently distinguish between tidally disturbed dwarfs
and tidal debris ejected during galaxy interactions (e.g., see the Appendix and
\citealt{Greco:2018aa}). To (at least slightly) reduce the impact of the
inevitable subjectivity in this process, we only keep sources that we both
flagged as a galaxy. 

In summary, we use \sextractor\ to build an initial catalog of
${\sim}7\times10^6$ sources.  We then make cuts based on size and color, as
measured by \sextractor, which reduces our sample to 14,069 objects. We model
each of these remaining sources using {\it imfit}, and we cut on the resulting
parameters to yield a sample of 1521 LSBG candidates. Finally, we visually
inspect the candidates, producing a final catalog of \ngal\ LSBGs, which will
be the focus of the remainder of the paper. 

\section{The Galaxy Sample}\label{sec:sample}

We have run our source-detection pipeline (Section~\ref{sec:pipeline}) on the
first ${\sim}$\area~deg$^{2}$ of the Wide layer of the \survey, which will grow
to 1400~deg$^{2}$ upon survey completion. The power of this survey is that it
is simultaneously wide, deep, and sharp, allowing for a homogeneous study of
LSB sources in environments ranging from the field to dense galaxy clusters.
Indeed, our search has uncovered a rich diversity of LSB phenomena, including
LSBGs, tidal debris from galaxy interactions, and scattered light from dust in
the Milky Way (Galactic cirrus). See the Appendix for representative examples
of the latter two types of LSB sources, which can be significant contaminants
in searches for LSBGs.

The primary goal of this work is to identify and characterize extended LSBGs.
Our final sample consists of \ngal\ galaxies, which span a wide range of
colors, morphologies, and environments. In this section, we separate our sample
into red and blue subsamples and present their observed properties.  \jpg{We
  publish our full LSBG catalog, the contents of which are summarized in
  Table~\ref{tab:catalog}, in machine-readable format.}
  
\subsection{Colors} \label{sec:colors}

\begin{figure}[t!]
  \includegraphics[width=\columnwidth]{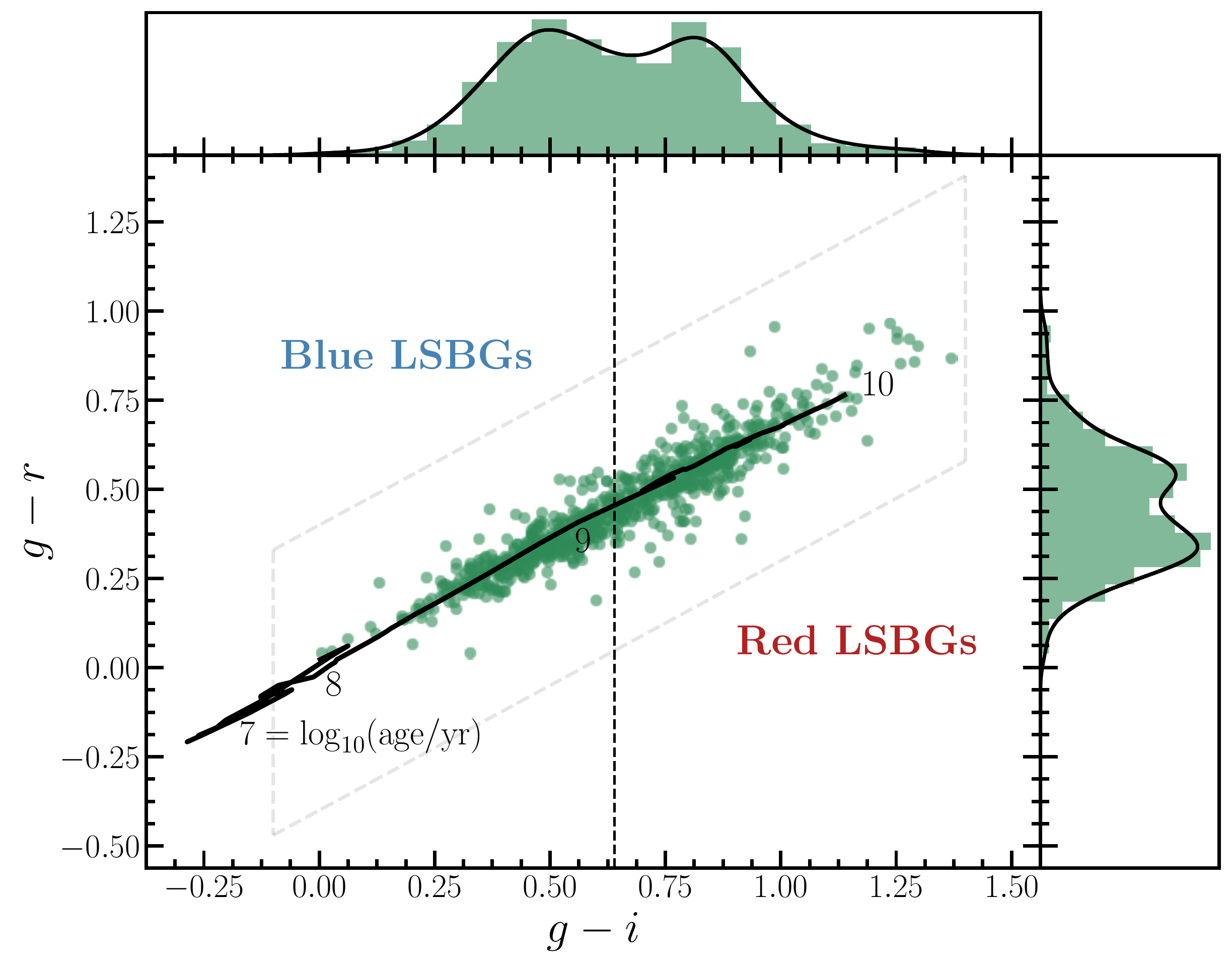}
  \caption{
    Color-color diagram for our full catalog of LSBGs. We separate the galaxies
    into red ($g-i\geq0.64$) and blue ($g-i<0.64$) subsamples, with the dividing
    color at the median value. We show the evolutionary path of a
    $0.4\times$solar metallicity simple stellar population from the models of
    \citet{Bruzual:2003aa}. The subsolar and solar metallicity models fall on
    very similar evolutionary paths in this color space. \resp{The black lines  
    overlaid on the histograms on the top and right side of the figure show the 
    density distributions obtained from kernel density estimation using a 
    Gaussian kernel. The dashed gray box shows our color selection region 
    (Section~\ref{sec:selection}).}}
    \label{fig:colors}
\end{figure}
\begin{figure*}[htbp]
  \includegraphics[width=\textwidth]{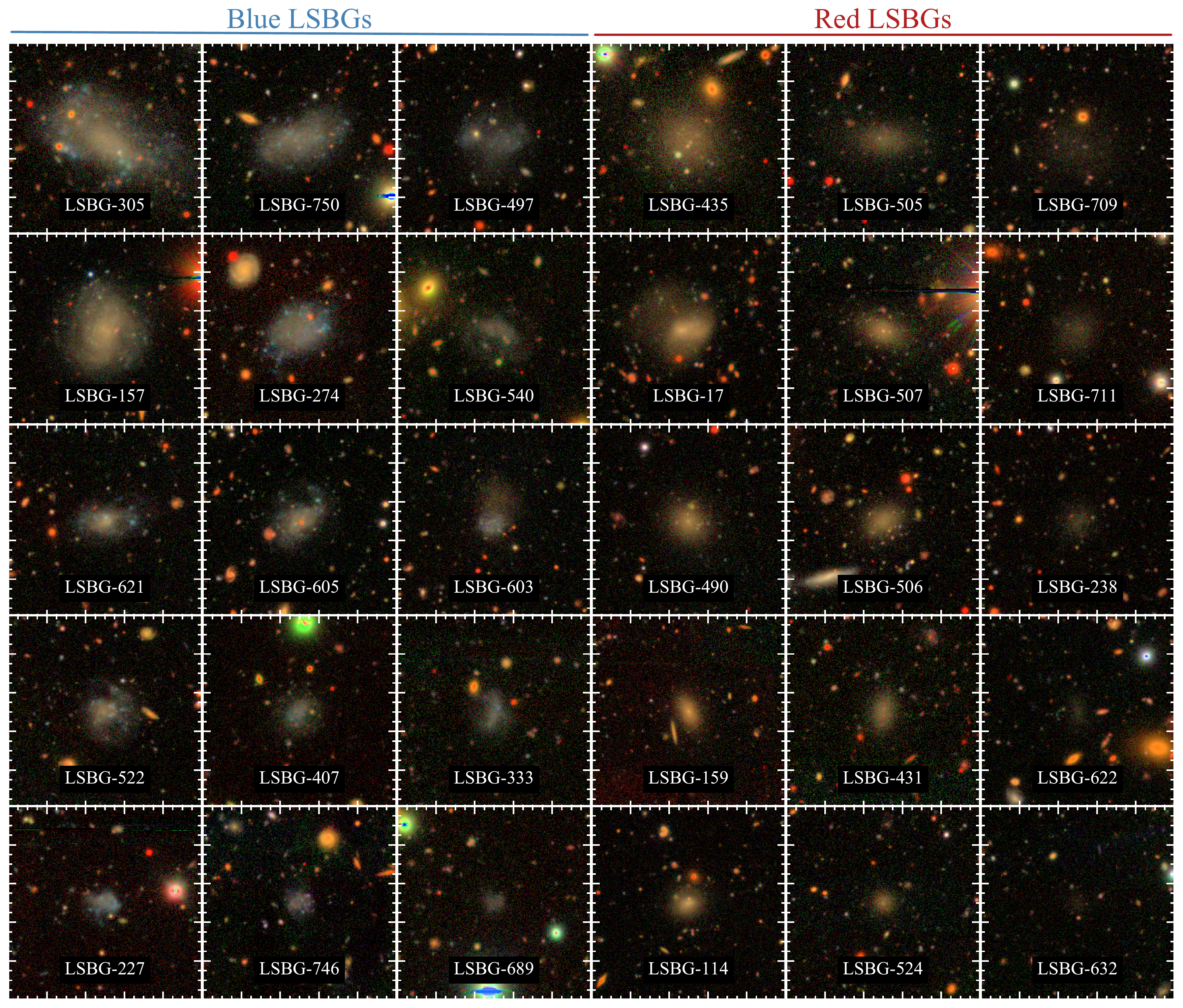}
  \caption{
    \survey\ $gri$-composite images \citep{Lupton:2004aa} of representative
    blue (left; $g-i<0.64$) and red (right; $g-i\geq0.64$) galaxies from our
    LSBG sample. Each cutout is 55\arcsec\ on a side. For each
    subsample, size (surface brightness) roughly deceases from top to bottom
    (left to right). The blue galaxies are generally irregular
    systems, whereas the red galaxies tend to be elliptical.
    \label{fig:stamps}
  } 
\end{figure*} 
\begin{figure*}[htbp]
  \includegraphics[width=\textwidth]{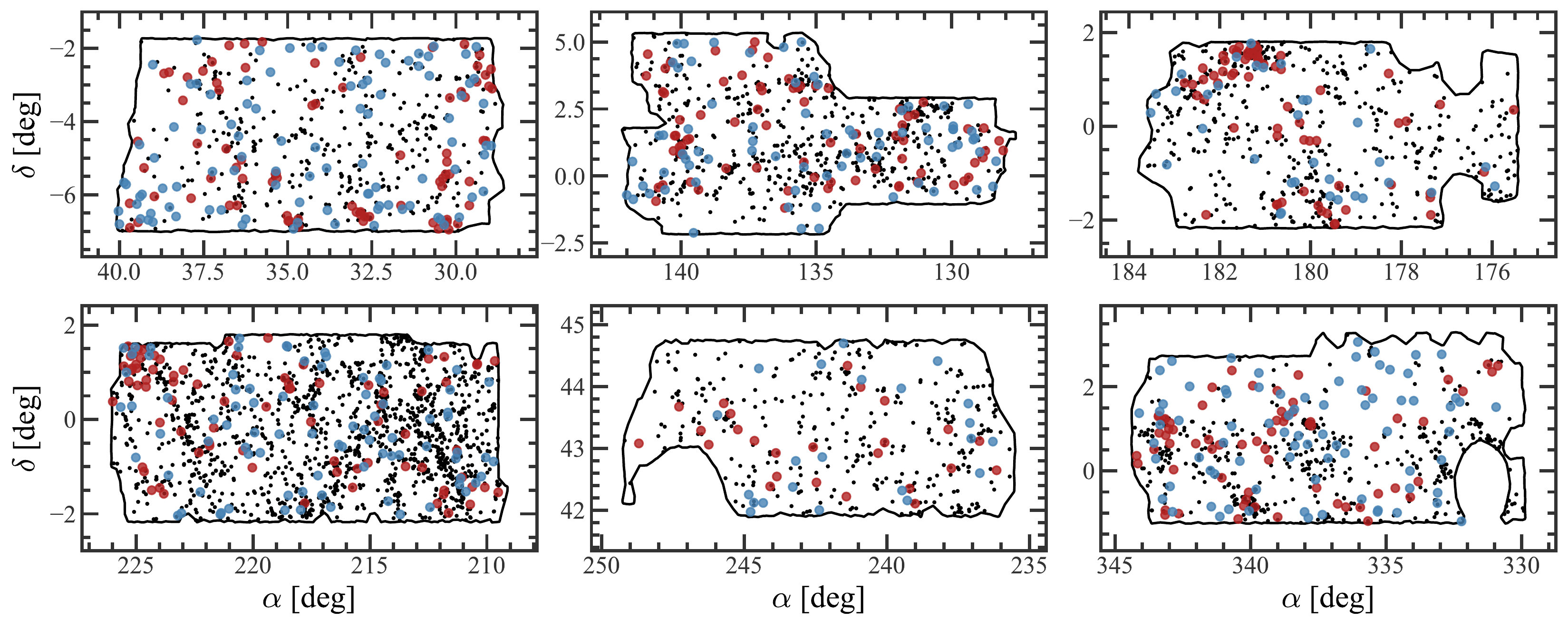}
  \caption{
    Sky positions of LSBGs within the six \survey\ fields that have been observed
    to the full Wide layer depth (in $gri$) as of the internal S16A data
    release (see \citealt{Aihara:2018ab} for information about the \survey\ data
    releases). Red LSBGs ($g-i\geq0.64$) are colored red, and blue LSBGs
    ($g-i<0.64$) are colored blue.  We also show the positions of galaxies with
    $z<0.055$ (black points) from the NASA-Sloan Atlas (NSA) galaxy catalog.
  }\label{fig:skypos} 
\end{figure*}

The optical colors of galaxies reveal their dominant stellar populations and
thus correlate strongly with galaxy morphology, leading to the well-known
separation of galaxies into red and blue sequences
\citep[e.g.,][]{Strateva:2001aa, Blanton:2009aa}. While large-scale surveys
such as SDSS have enabled detailed studies of the color distribution of
galaxies down to $\bar{\mu}_\mathrm{eff}(r)\sim24.5$~\sbunit\
\citep[e.g.,][]{Baldry:2004aa}, less is known about the color distribution at
lower surface brightnesses. Classical disk LSBGs are known to span blue to red
colors \citep{ONeil:1997aa}. Blue colors are generally associated with spiral
or irregular systems, whereas red colors are indicative of spheroids or
ellipticals \citep{Mateo:1998aa}, with quenched galaxies being found almost
exclusively in association with massive host systems \citep{Geha:2012aa}. For
present-day, red UDGs in groups/clusters, there is evidence that suggests they
formed as (physically large) blue LSB dwarf galaxies in the field and were
transformed during their infall onto dense environments \citep{Roman:2017ab}.
Motivated by these previously observed trends, we start by presenting the
colors of our sample and defining blue and red galaxies, which will be a useful
point of comparison.

Intriguingly, our LSBG sample shows evidence for color bimodality in both $g-r$
and $g-i$, with a clear correlation between color and galaxy morphology.  In
Figure~\ref{fig:colors}, we show our sample in the color-color diagram $g-r$
vs. $g-i$. We separate red and blue galaxies using the median $g-i$ color
(dashed black line): red galaxies are defined to have $g-i\geq0.64$ and blue
galaxies are defined to have $g-i<0.64$. We note that modeling the $g-i$ color
distribution as a sum of two Gaussians yields a color boundary within
0.01~mag of the median; we choose to use the median for simplicity. Evidence
for color bimodality can be seen in the associated histograms on the top and
right side of the figure. \resp{As an additional test of the apparent 
bimodality, we performed kernel density estimation on both distributions, 
assuming Gaussian kernels with bandwidths determined by a cross-validation 
grid search over a reasonable range of widths. 
The results are shown as black lines overlaid on each histogram; we see 
that the bimodality in both color distributions appears to be robust to 
our choice of density estimation method.}

\resp{Our sample's median $g-i$ color of 0.64 is consistent with UDG candidates  
around Abell Cluster 168 and its surrounding large-scale structure 
\citep{Roman:2017aa}; this galaxy sample is composed of blue and 
red LSBGs with median $g-i = 0.66$. The red LSBGs in our sample have 
median $g-i = 0.8$, which is similar to the population of UDGs in the Coma 
cluster \citep{van-Dokkum:2015aa}. In contrast, our blue LSBGs are much bluer  
with median $g-i=0.47$, which is comparable to the average $g-i$ color of 0.45
observed in a sample of \HI-rich UDGs in the field \citep{Leisman:2017aa}.}

In Figure~\ref{fig:colors}, we also show the evolutionary path of a
$0.4\times$solar metallicity stellar population from the models of
\citet{Bruzual:2003aa}. The subsolar and solar metallicity models follow very
similar tracks in this color space, so we cannot set meaningful metallicity
constraints with the available colors. The red LSBGs are consistent with being
composed of mostly old stars, whereas the colors of the blue LSBGs require
young stellar populations. 
 
In addition to their difference in color, the red and blue galaxies have
visibly distinct morphologies. In Figure~\ref{fig:stamps}, we show
$gri$-composite images of representative galaxies from each subsample, spanning
the distribution of surface brightnesses and sizes. The blue galaxies tend to
have irregular, lumpy morphologies and compact star-formation regions.  In
addition, most (76\%) of the blue sources have ultraviolet (UV) detections in
the GALEX source catalog (see Section~\ref{sec:galex}), suggesting ongoing star
formation in these systems. In contrast, the red galaxies typically have
spherical or elliptical shapes with light profiles that are well-characterized
by single-component S\'{e}rsic functions. The observed properties of these
galaxies are consistent with early-type dwarf galaxies, and some fraction may
be physically similar to UDGs in dense galaxy clusters, depending upon their
(unknown) distances.

In the following sections, we present and compare properties of the red and
blue subsamples following the above $g-i$ color definition. 

\begin{figure*}[t!]
  \includegraphics[width=\textwidth]{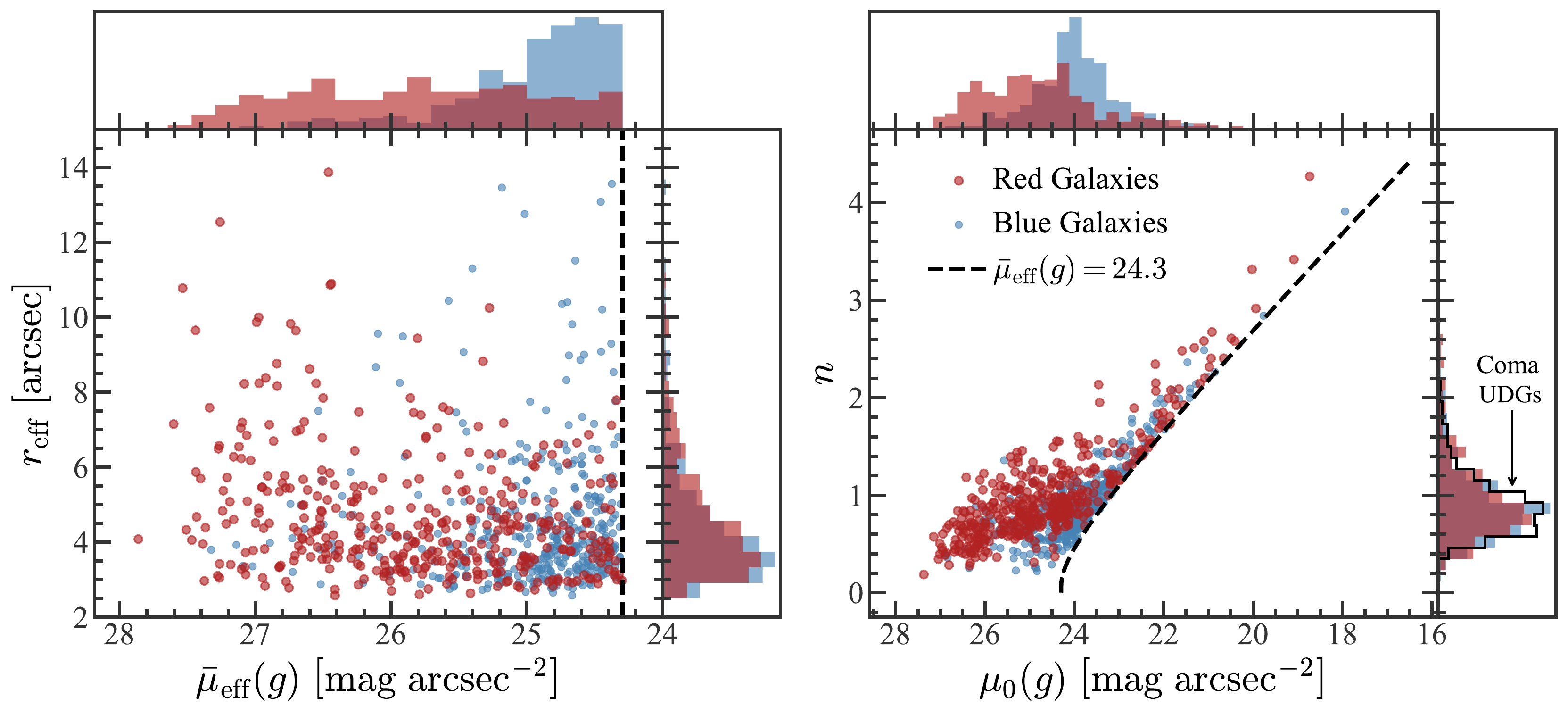}
  \caption{
    {\it Left:} Relationship between effective radius $r_\mathrm{eff}$ and
    $g$-band mean surface brightness within the circularized effective radius
    \mueff\ for our LSBG sample.  The independent parameter distributions are
    shown as histograms on the top and right side of the figure. Points and
    histograms are colored according to the galaxy color definition described
    in Section~\ref{sec:colors}. {\it Right:} S\'{e}rsic index $n$ vs. $g$-band
    central surface brightness $\mu_0(g)$ for the same galaxy samples shown in
    the left panel. Since we select galaxies based on \mueff, there is a
    high-surface-brightness tail associated with high S\'{e}rsic $n$ values.
    These high-$n$ (and thus high-$\mu_0$) galaxies, which make up less than
    5\% of our sample, generally have bright cores or central star-forming
    regions. For comparison, we show the S\'{e}rsic $n$ distribution of UDGs in
    the Coma cluster \citep{Yagi:2016aa} as the black line in the horizontal
    histogram on the right. \jpg{In both panels, the dashed black line shows
      our selection cut at \mueff$=24.3$~\sbunit, which results in a curve in
      the $\mu_0-n$ plane.
    }
  }\label{fig:mu} 
\end{figure*} 

\begin{figure}[htbp]
  \includegraphics[width=\columnwidth]{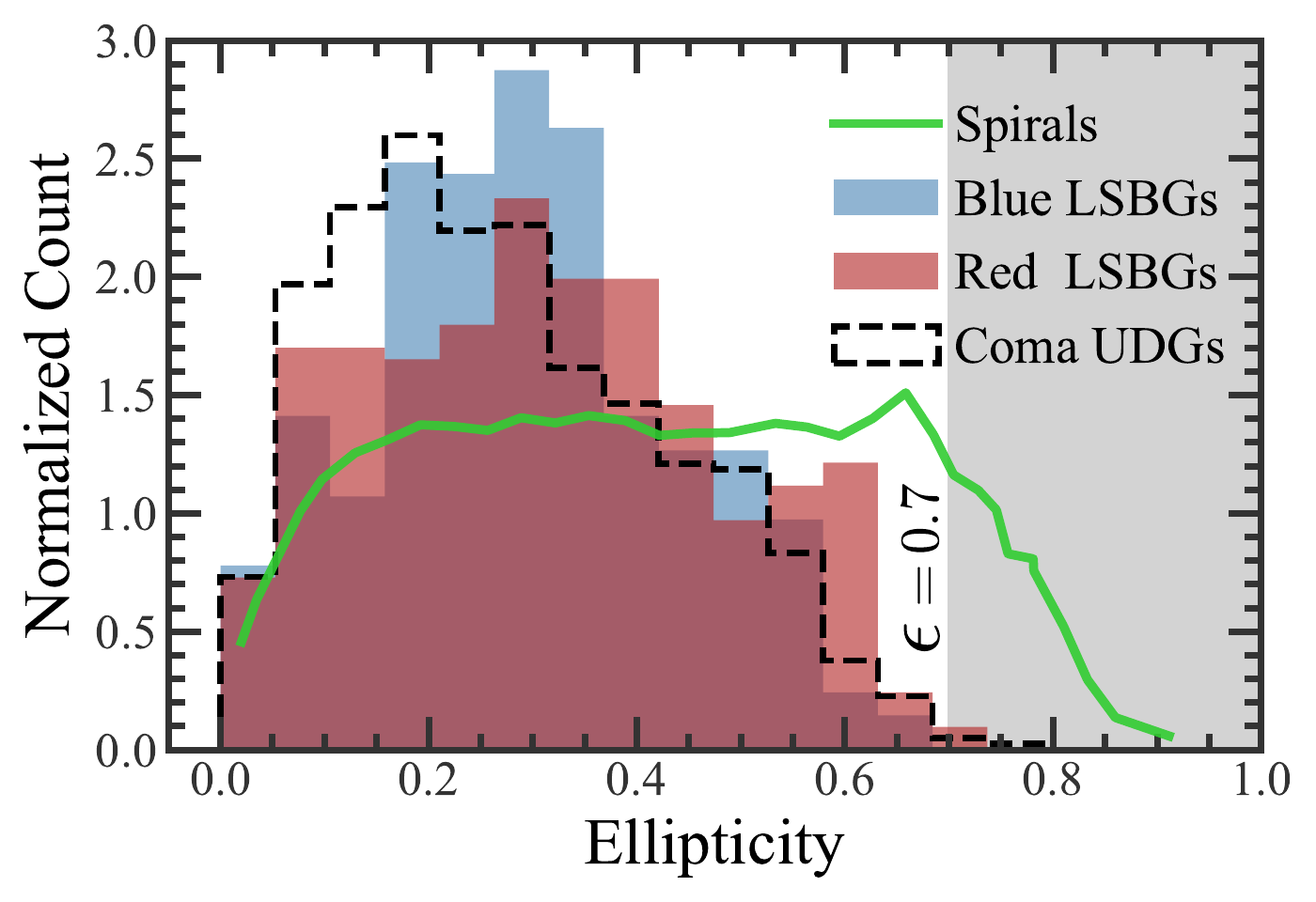} \caption{ Ellipticity
    ($\epsilon$) distributions of our red and blue LSBG subsamples, UDGs in the
    Coma cluster \citep{Yagi:2016aa}, and spiral galaxies from SDSS
    \citep{Rodriguez:2013aa} that were classified by the Galaxy Zoo project
    \citep{Lintott:2008aa}. \jpg{We select galaxies with $\epsilon<0.7$; values
      outside this range are shaded gray.} } \label{fig:ell} 
\end{figure}

\subsection{Spatial Distribution}\label{sec:spatial}

In Figure~\ref{fig:skypos}, we show the sky positions of LSBGs in our sample
(colored points) within each \survey\ field. Following the above definition,
red LSBGs are shown as red points and blue LSBGs as blue points.  Our selection
on large size ($r_\mathrm{eff}>2.5\arcsec$) likely restricts most of our sample
to low redshift. For reference, we plot the positions of low-$z$ galaxies with
known redshifts from the NASA-Sloan
Atlas\footnote{\url{http://nsatlas.org}} (NSA), which contains virtually all
galaxies with known redshifts out to $z = 0.055$ within the coverage of SDSS
Data Release (DR) 8 \citep{Aihara:2011aa}. The overlap between the NSA 
and \survey\ is not perfect, so regions with few black points 
(e.g., the northern region in the bottom-right panel) are not necessarily 
representative of the associated low-$z$ galaxy population.

Without spectroscopic redshifts for a large fraction of the objects in our
sample, a cross-correlation analysis \citep[e.g.,][]{Menard:2013aa} may be used
to estimate the redshift distribution and statistically infer the physical
nature of these objects. We leave such an analysis for future work.
\jpg{Interestingly, however, we can already see that sources in our sample are not
distributed uniformly on the sky---many cluster both with other LSBGs
and with higher-surface-brightness low-$z$ galaxies, and the
effect appears to be stronger for the red LSBGs (see Section~\ref{sec:ngc-group}
for a detailed look at one of these overdensities).
However, it is important to emphasize that detailed auto- and cross-correlation
analyses are necessary to make robust claims about the clustering of these 
objects.}

\begin{figure*}[t!]
  \includegraphics[width=\textwidth]{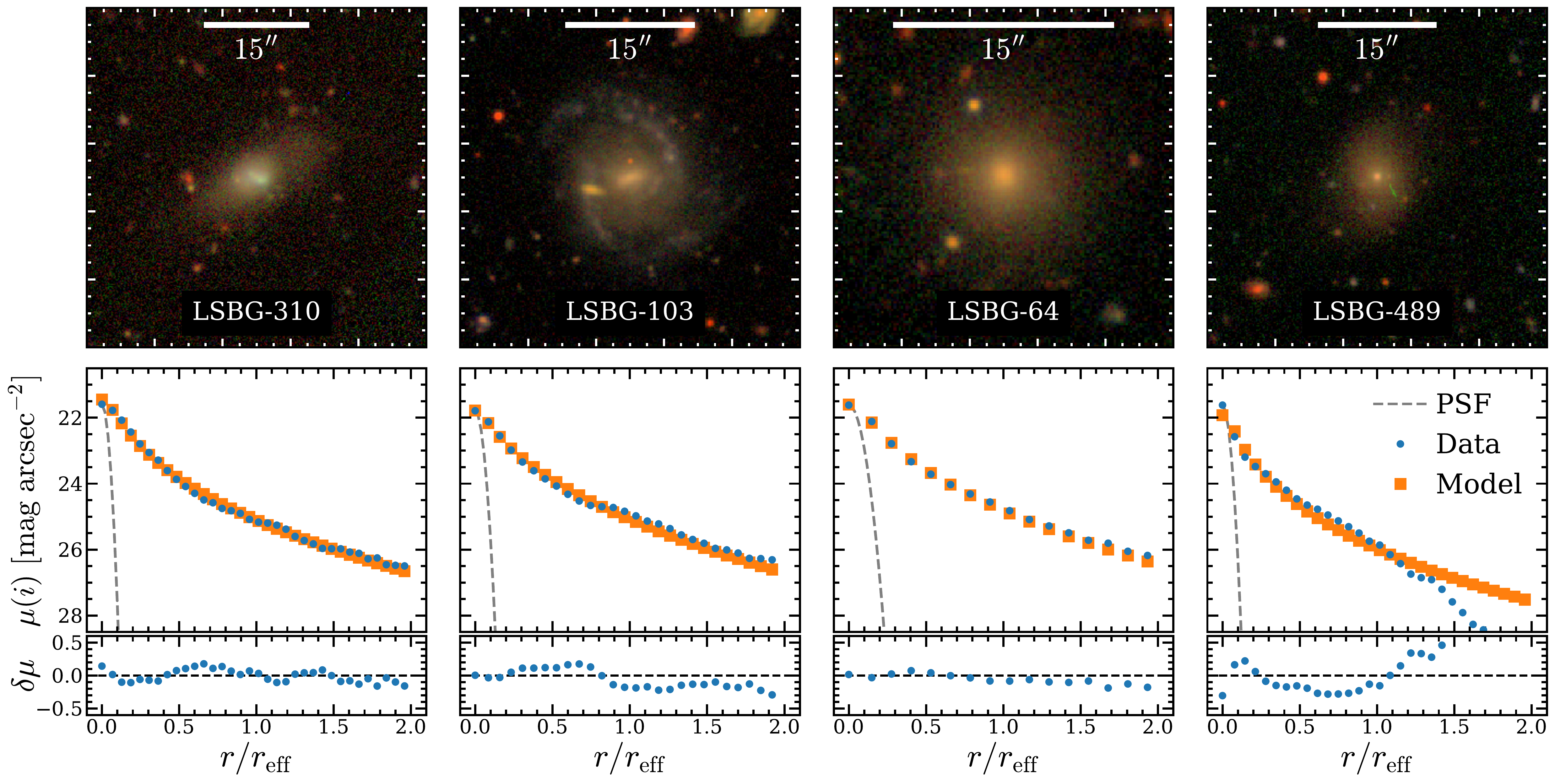}
  \caption{\resp{{\it Top:} Example $gri$-composite \survey\ images of high-$\mu_0$
  (and thus high-$n$) sources, which represent $<5$\% of our sample (see right 
  panel of Figure~\ref{fig:mu}). From left to right, we show galaxies whose 
  central surface brightness is high due to a central star forming region, an
  apparent bulge within a LSB disk, a bulge-like core in a red spheroidal galaxy, and 
  an apparent point source (nuclear cluster). The angular scale is indicated by
  the white bar at the top of each panel. {\it Bottom:} Radial $i$-band 
  surface-brightness profiles of each galaxy (blue dots), where we use elliptical 
  apertures with shape parameters given by our two-dimensional S\'{e}rsic 
  fits. For each galaxy, we scale the radial coordinate by the effective radius. 
  For comparison, we show the angular scale of the PSF (gray dashed lines) and 
  model profiles generated using the same apertures on two-dimensional, PSF-convolved
  S\'{e}rsic model images (orange squares). The lower panels show the residuals 
  between the measured profiles and the S\'{e}rsic models.}}
  \label{fig:high-mu0-sources} 
\end{figure*} 

\subsection{Surface Brightness and Shape Distributions}

We now turn to the distributions of the best-fit S\'{e}rsic parameters. In the
left panel of Figure~\ref{fig:mu}, we compare the relationship between
effective radius $r_\mathrm{eff}$ (measured along the major axis) and mean
surface brightness \mueff\ (measured within the {\it circularized} effective
radius) for the red and blue subsamples.  The radius distributions are very
similar --- both span a wide range of $r_\mathrm{eff}\sim2.5$-$14$\arcsec\ and
have a median value of ${\sim}4\arcsec$. In contrast, the mean surface
brightness distributions are dramatically different. The red distribution is
much broader with 16th, 50th, and 84th percentiles of \mueff\,$=24.8$, 25.8,
and 26.8~\sbunit, respectively.  For the blue galaxies, the same percentiles
are \mueff\,$=24.5$, 24.8, 25.5~\sbunit.  Hence, the blue LSBGs typically 
have higher mean surface brightnesses than the red LSBGs. 
In addition, galaxies that are both faint 
$\left( \bar{\mu}_\mathrm{eff}(g)> 26~\mathrm{mag\ arcsec^{-2}}\right)$ and 
large in angular extent $\left(r_\mathrm{eff}>6\arcsec\right)$ are almost 
exclusively red. \jpg{The black dashed line shows our selection 
cut at \mueff$= 24.3$~\sbunit.}

In the right panel of Figure~\ref{fig:mu}, we show S\'{e}rsic index $n$ vs.
central surface brightness $\mu_0(g)$. In general, the blue LSBGs 
have brighter central surface brightnesses than the red LSBGs, with median 
values of $\mu_0(g) = 24.0$~\sbunit\ and $\mu_0(g) = 24.9$~\sbunit, 
respectively. The red distribution is again the broader of the two---79 
red galaxies have $\mu_0(g)>26$~\sbunit, whereas only 11 blue galaxies 
reach such low central surface brightnesses. Again, the black dashed line 
represents our cut on \mueff, which produces a curve in the $\mu_0-n$ plane. 
The horizontal histogram in the right panel of Figure~\ref{fig:mu} compares the
red and blue S\'{e}rsic index distributions. For comparison, the black
line shows the same distribution for a sample of 753 UDGs in the Coma cluster
\citep{Yagi:2016aa}. All three distributions favor nearly exponential light
profiles, with a median index of $n\sim0.9$.  

We show ellipticity distributions in Figure~\ref{fig:ell}. The LSBGs are
generally round, with a median $\epsilon = 0.3$ for the full sample. While our
selection removes sources with $\epsilon>0.7$, the distribution begins to fall
before this threshold, particularly for the blue galaxies. The dashed black
line shows the distribution for UDGs in Coma, which are also overwhelmingly
round systems. For randomly oriented disks, the expected ellipticity
distribution is much flatter between $\epsilon = 0.1$-0.7. This can be seen in
the green line, which shows the distribution for a sample of spiral galaxies
from SDSS \citep{Rodriguez:2013aa} that were classified by the Galaxy Zoo
project \citep{Lintott:2008aa}.

\subsection{High-$\mu_0$ Sources}\label{sec:high-mu0}

Both the red and blue $\mu_0(g)$ distributions have a high-surface-brightness
tail, comprising a very small fraction of the sample.
Less than 5\% of our sample has $\mu_0(g)<22$~\sbunit. This is due to our
selection on \mueff\ rather than $\mu_0(g)$. As described in
Section~\ref{sec:modeling}, we cut on mean surface brightness to allow
nucleated sources into our sample.  Consistent with our selection, high central
surface brightnesses are associated with high $n$ values, which are indicative
of compact, relatively bright cores. In most cases 
\resp{(${\sim}50$\% of high-$\mu_0$ sources)}, these cores are 
bright \resp{bulge-like} components embedded in an LSB disk, 
similar to what is seen in the population of giant LSB spirals 
\citep[e.g.,][]{Sprayberry:1995aa} of which Malin 1 \citep{Bothun:1987aa} 
represents the most extreme example. \resp{In the other cases, there is a bright 
central star formation region or potential nuclear cluster that drives the 
fit toward high-$n$/high-$\mu_0$ values.} 
  
\resp{In the top row of Figure~\ref{fig:high-mu0-sources}, we show 
$gri$-composite images of example high-$\mu_0$ sources. From left to right, 
the central surface brightness is high due to a central star 
formation region, a bulge within a LSB spiral, a bulge-like core within 
a smooth red spheroid (in projection), and an apparent central point 
source (nuclear cluster). The bottom row of this figure shows the 
surface brightness profile of each galaxy extracted within elliptical 
apertures using the shape parameters from \code{imfit} 
(Section~\ref{sec:modeling}). We also show profiles extracted using the 
same apertures on two-dimensional model images, where each 
galaxy is modeled as a PSF-convolved S\'{e}rsic function with the best-fit 
parameters from \code{imfit}. The central surface brightnesses---which, in our 
catalog, are based on the deconvolved S\'{e}rsic model---are higher 
than traditional definitions of LSBGs (optical $\mu_0$ fainter than 
${\sim}22$-23~\sbunit). However, these sources are low surface 
brightness on average, and their relation to non-nucleated/bulgy LSBGs deserves 
further investigation. We therefore include them in our catalog, noting 
that they represent a small fraction of our sample.}

\section{Source Crossmatching} \label{sec:xmatch}

Given the LSB nature of our catalog, few sources have previous redshift
measurements, and most archival photometric data lack the depth needed to
accurately characterize our sources. Nonetheless, it is still possible to gain
insight into the physical properties spanned by galaxies in our sample by
crossmatching with existing catalogs. Therefore, in this section, we present
results from crossmatching with archival data ranging from radio to UV
wavelengths. We show LSBGs with distance information in the size-luminosity
plane in Section~\ref{sec:size-lum}.

\subsection{External Wide-Field Surveys}

\subsubsection{SDSS Photometric Catalog}\label{sec:sdss}

\begin{figure}[t!]
  \includegraphics[width=\columnwidth]{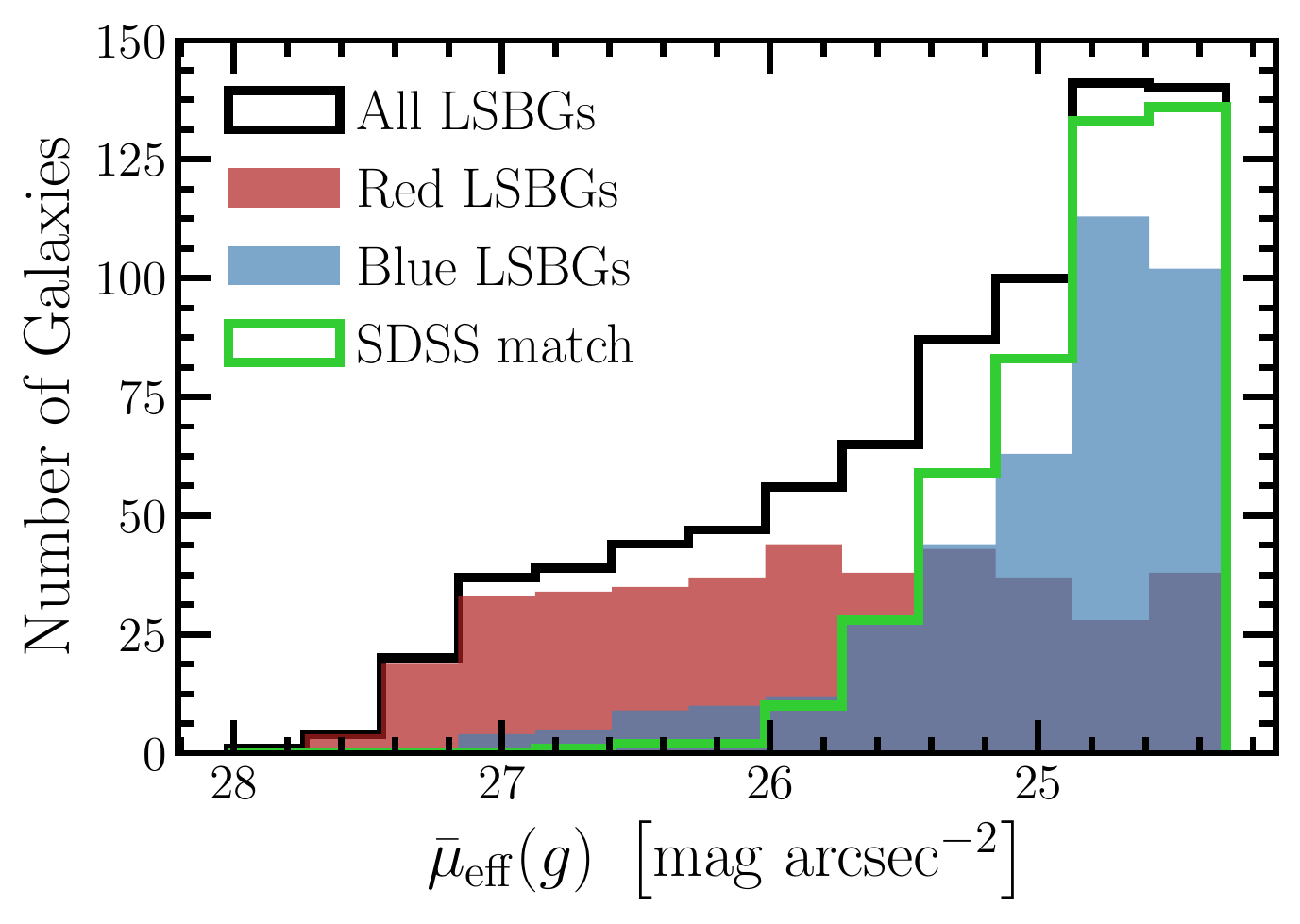}
  \caption{
    Distribution of mean surface brightness for LSBGs with a detection in the
    SDSS DR12 photometric catalog (green histogram), where we consider objects
    within 3\arcsec\ a match. We also show the distribution for our full galaxy
    sample (black histogram) and our blue (blue histogram) and red (red
    histogram) subsamples. Note that the SDSS detections are near the detection 
    limit of the survey, making the photometry less reliable 
    (see Figure~\ref{fig:galex} for SDSS images of LSBGs). 
    }
    \label{fig:sdss}
\end{figure}

\begin{figure*}
  \centering
  \includegraphics[width=\textwidth]{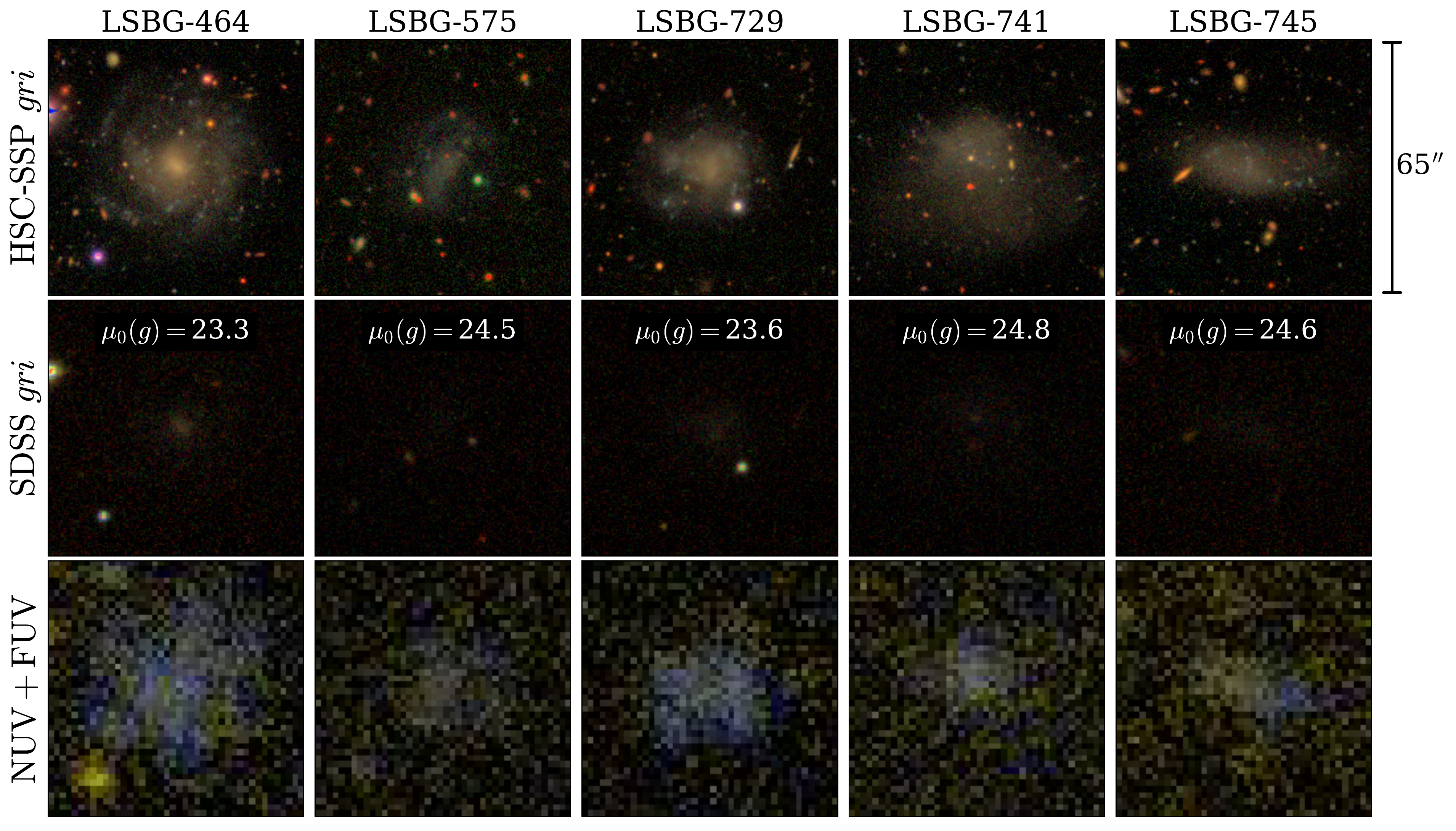}
  \caption{
    Cutout images from \survey\ (top row), SDSS (middle row), and GALEX (bottom
    row) showing some of the largest LSBGs in our sample. The $gri$-composite
    images from \survey\ and SDSS were created using the same stretch following
    the method of \citet{Lupton:2004aa}. Each cutout image is 65\arcsec\ on a
    side.  LSBG-464 ($z=0.02573$) and LSBG-575 (distance = 29.5~Mpc) are in the
    70\% ALFALFA catalog (see Section~\ref{sec:alfalfa}), and LSBG-729
    ($z=0.02513$) hosted a Type IIb supernova in 2009 (see
    Section~\ref{sec:SN-host}). 
  } 
  \label{fig:galex} 
\end{figure*}

While the vast majority of sources in our sample have surface brightnesses near
or below the detection limit of SDSS \citep[e.g.,][]{Blanton:2005aa}, many
still have detections in the SDSS photometric catalog. In
Figure~\ref{fig:sdss}, we show the distribution of \mueff\ 
(as we measured with \survey) for galaxies with a
matched source in SDSS DR12, where we consider objects within 3\arcsec\ a
match. The SDSS detections drop quickly to zero for surface brightnesses
fainter than 25.5~\sbunit, and as the red histogram shows, more than half of
our red LSBGs exist within this region of parameter space. At these surface
brightnesses, it is important to note that a detection in the SDSS photometric
catalog does not necessarily mean the photometry is reliable.  For example, if
the center of an LSBG is detected, its outer profile may fall below the
detection limit, resulting in its size and magnitude being underestimated by
the SDSS photometric pipeline. See the middle row of Figure~\ref{fig:galex} for
example SDSS $gri$-composite images of some of our largest LSBGs.


\subsubsection{The NASA-Sloan Atlas}\label{sec:nsa}

Redshifts are crucial for interpreting the physical nature of individual
galaxies. Unfortunately, our LSBGs are generally too faint to be in the SDSS
spectroscopic catalog (two exceptions are given below). To search for LSBGs
with previous (optical) spectroscopic redshifts, we crossmatch with the NSA
galaxy sample, which was selected to include virtually all known redshifts 
out to $z=0.055$ within the coverage of SDSS DR8. We find three 
matches that are within 2\arcsec\ from objects in our sample.

Two of the matches are physically large, face-on LSB spirals from the SDSS
spectroscopic catalog: LSBG-171 (NSA ID 42601) and LSBG-456 (NSA ID 145288).
LSBG-171 is at $z=0.04389$ and has $r_\mathrm{eff}=6.7$~kpc and
$\mu_0(g)=22.1$~\sbunit\ (we correct for cosmological dimming when redshift
information is available). LSBG-456 is at $z=0.02863$ and has
$r_\mathrm{eff}=5.9$~kpc and $\mu_0(g)=23.8$~\sbunit, similar in size and
central surface brightness to UDGs, although its absolute magnitude of
$M_g=-17.9$ is nearly 3 magnitudes brighter than the most luminous UDGs
discovered by \citet{van-Dokkum:2015aa}.

The third match is LSBG-613 (NSA ID 144517). This LSBG has a redshift of
$z=0.02447$, which was measured during a survey for LSB dwarf galaxies
\citep{Roberts:2004aa}. Its inferred physical effective radius is
$r_\mathrm{eff}=3.3$~kpc, and its central surface brightness is
$\mu_0(g)=23.9$~\sbunit, again putting it near the UDG size-surface brightness
parameter space.

\subsubsection{ALFALFA}\label{sec:alfalfa}

The gas mass fractions of LSB dwarf galaxies are among the highest of any known
galaxy type (as high as $f_\mathrm{gas}=0.95$; \citealt{Schombert:2001aa}).
Such gas-rich systems are probed by radio surveys tuned to measure the 21~cm
line of atomic hydrogen (\HI). The largest volume blind \HI\ survey to date is
the Arecibo Legacy Fast Arecibo L-band Feed Array
\citep[ALFALFA;][]{Giovanelli:2005aa}. We crossmatch our sample with the 70\%
ALFALFA catalog\footnote{The ALFALFA 70\% catalog is publicly available at
  \url{http://egg.astro.cornell.edu/alfalfa/data/index.php}.}
\citep{Haynes:2011aa}, which covers 70\% of their final survey area; 361 of our
LSBGs fall within this footprint.

The ALFALFA team visually inspected every \HI\ source in their catalog and
searched for optical counterparts using optical images from the Palomar Digital
Sky Survey and, where available, the SDSS. We crossmatch our sample with their
assigned optical coordinates and find 3 matched sources. Given the higher image
quality afforded by \survey, we also crossmatch with ALFALFA's full list of
\HI\ coordinates to crosscheck their chosen optical counterparts. We consider
any source with \HI\ coordinates within 1$\farcm$75 (approximately half of
Arecibo's beam) of an LSBG in our sample. This produced 17 matches with an
ALFALFA detection code of 1 (12 sources) or 2 (5 sources), where code 1 refers to 
high signal-to-noise ratio \HI\ detections with reliable optical 
counterparts, and code 2 refers to low signal-to-noise ratio \HI\ detections
with optical counterparts whose (optical) redshifts are consistent with 
the \HI\ line measurements \citep{Haynes:2011aa}. We then visually inspected 
each match. We recover the 3 optical matches above, and in all other
cases, the optical counterpart assigned by the ALFALFA team is a
higher-surface-brightness galaxy.  While we cannot rule out the possibility
that any of our matched sources are also contributing to the observed \HI\
signals, we assume the ALFALFA-selected optical counterparts are correct.
Hence, we have 3 LSBGs with an \HI\ counterpart in the 70\% ALFALFA catalog.  

Two of the matches are physically large face-on LSB spirals, one of which was
matched with the NSA above: LSBG-456 (NSA ID 145288 and AGC 243463) and LSBG-464
(AGC 249425). Similar to the NSA-matched galaxies, LSBG-464's redshift of
$z=0.02573$ implies it is physically large, with an effective radius of
$r_\mathrm{eff}=6.0$~kpc. Images of LSBG-464 from \survey, SDSS, and GALEX can
be seen in the first column of Figure~\ref{fig:galex}. While there is a clear
detection of the galaxy's core in the SDSS image, the giant disk and spiral
arms are only visible in the \survey\ image, highlighting the need for deep
imaging in searches for LSBGs. The strong detection of the disk in the UV by
GALEX reflects ongoing star formation throughout the disk. 

The third match with ALFALFA is LSBG-575 (AGC 189086). We show images of this
source from \survey, SDSS, and GALEX in the second column of
Figure~\ref{fig:galex}. This LSBG is relatively nearby at a distance of
29.5~Mpc, its \HI\ mass is $\log_{10}(M_\mathrm{HI}/M_\odot)=8.08$, and it has
a low velocity width (at 50\% of the peak flux) of 26~km~s$^{-1}$. For context,
the mean width for all ALFALFA galaxies is 194~km~s$^{-1}$
\citep{Leisman:2017aa}. From \survey\ optical data, we find that LSBG-575 has a
central surface brightness of $\mu_0(g)\sim24.5$~\sbunit. Its irregular
morphology makes size measurements highly uncertain; nevertheless, we estimate
it to have a large effective radius of $r_\mathrm{eff}=11\arcsec$, which at its
distance corresponds to a physical size of $r_\mathrm{eff}=1.6$~kpc. These
physical properties are consistent with LSBG-575 being a gas-rich UDG.

Assuming the mass-to-light ratio/color relation derived from
\citet{Bell:2003aa}, LSBG-575 has a stellar mass of
$M_\star\sim2.6\times10^7~M_\odot$, implying an \HI\ to stellar mass ratio of
$M_\mathrm{HI}/M_\star = 4.7$, a factor of 7 lower than the mean value of the
\HI-bearing UDGs discovered by \citet{Leisman:2017aa}.  Following
\citet{Schombert:2001aa}, we use this ratio to estimate the baryonic gas
fraction as 
\begin{equation}\label{eqn:f_gas}
  f_\mathrm{gas} = \frac{M_\mathrm{gas}}{M_\mathrm{gas} + M_\star} \approx 
                   \left(1 + \frac{M_\star}{1.4\,M_\mathrm{HI}}\right)^{-1}, 
\end{equation} 
where $M_\mathrm{gas}$ is the total gas mass, and the factor of 1.4 assumes a
solar hydrogen mass fraction. For LSBG-575, we find a high gas fraction of
$f_\mathrm{gas}\approx0.87$, which is not uncommon at this stellar mass
\citep[e.g.,][]{Geha:2006aa,Kim:2007aa,Bradford:2015aa}.

In summary, LSBG-575 has a low \HI\ mass compared to the ALFALFA population
(but consistent with the expectation given its stellar mass) and a high gas
fraction, which is typical for objects of its stellar mass but lower than other
HI-bearing UDGs \citep[e.g.,][]{Papastergis:2017aa}.

\subsubsection{GALEX}\label{sec:galex}

A galaxy's integrated UV light, which is dominated by young, massive stars, is
a well-known tracer of its current star formation rate (SFR). To get a sense
for the fraction of objects in our sample with ongoing star formation, we
crossmatch with the GALEX source catalog \citep{Martin:2005aa} using the MAST
Portal\footnote{\url{https://mast.stsci.edu/}}.  We adopt a separation
threshold of 4\arcsec\ \citep[e.g.,][]{Budavari:2009aa} for matched sources.
Following the GALEX DR6 documentation, we drop matched sources with
artifact flags associated with dichroic reflections (\textsc{nuv\!\_artifact =
  4} or \textsc{fuv\!\_artifact = 4}) and/or window reflections
(\textsc{nuv\!\_artifact = 2}; NUV detector only). In addition, we drop sources
with negative NUV magnitudes. Since the MAST archive contains all GALEX
observations, sources with repeated observations have multiple entries in the
catalog. For each matched source, we keep the GALEX observation  with the
longest NUV+FUV exposure time. If multiple observations have the same exposure
time, we keep the source with the smallest offset from the position of the
matched LSBG in our catalog. 

Most of the sources in our catalog (761 out of 781) fall within the footprint
of at least one GALEX survey\footnote{To determine which sources were covered
  by any GALEX pointing, we crossmatched our LSBG catalog with the centers of
  all individual GALEX visits and checked if the positions fell within the
  field-of-view radius from the center of any observation, as suggested by
  \citet{Bianchi:2017aa}.}. Of these GALEX-observed LSBGs, 374 (${\sim}50\%$)
have UV detections. Of these detections, 80\% are blue LSBGs, and of all the
blue LSBGs in our sample, 76\% have a UV counterpart in GALEX. The red LSBGs
with GALEX detections (78 in total) may be interesting objects for follow-up
studies---about half have $g-i>0.7$.

\jpg{For convenience, we give the existing NUV and FUV magnitudes for all
  matched sources, as well as the GALEX survey from which the data were taken,
  in machine-readable format; see Table~\ref{tab:catalog} for a summary of the
  contents of the catalog.} We simply adopt the GALEX catalog magnitudes; these
are not matched aperture measurements.  Note that the GALEX surveys (i.e., the
All-Sky Imaging, Medium Imaging, Deep Imaging, and Nearby Galaxy Surveys, and
the Guest Investigator Program) image the sky to different depths.  

In Figure~\ref{fig:galex}, we show cutout images from \survey, SDSS, and GALEX
of some the largest LSBGs in our sample that also have GALEX detections. We
show the SDSS images to give the reader a context from which to view the
\survey\ images, which push to much lower surface brightness levels than is
commonly seen from wide-field surveys. 

The LSBGs with known distances suggest that our sample spans a distance
range of (at least) ${\sim}30$-100~Mpc. Assuming this distance range, we
estimate the range of SFRs as \citep{Kennicutt:1998ab}: 
\begin{equation}\label{eqn:sfr}
  \mathrm{SFR}\ [M_\odot\ \mathrm{yr^{-1}}] = 
  1.4\times10^{-28}\,L_\nu\ [\mathrm{erg\ s^{-s}\ Hz^{-1}}], 
\end{equation}
where $L_\nu$ is the UV luminosity derived from the extinction-corrected FUV
magnitudes. We estimate the Galactic reddening $E(B-V)$ using the maps of
\citet{Schlegel:1998aa} and the \citep{Cardelli:1989aa} extinction law with
$R_V = A_V/E(B-V) = 3.1$ and $A_\mathrm{FUV}=8.24\,E(B-V)$
\citep{Wyder:2007aa}; we do not account for internal dust extinction.  For the
assumed distance range, the median SFR of our sample (including only objects
with FUV measurements) spans a range of
${\sim}0.002$-$0.02~M_\odot~\mathrm{yr^{-1}}$, where these values are likely lower
limits, since we have ignored internal dust extinction.  Assuming the
mass-to-light ratio/color relation derived from \citet{Bell:2003aa}, this same
distance range implies a median stellar-mass range of
$M_\star\sim10^7$-$10^8~M_\odot$. With such parameters, these objects are
consistent with the observed SFR--$M_\star$ relation for \HI-selected galaxies
from the ALFALFA survey \citep{Huang:2012ab,Leisman:2017aa}. 

\subsection{Early-type dwarfs in a nearby galaxy group}\label{sec:ngc-group}

\jpg{As noted in Section~\ref{sec:spatial}, many of the red LSBGs appear to
  cluster with one another, as well as with higher-surface-brightness low-$z$
  galaxies.} This clustering can be used to statistically estimate the
distances to an ensemble of sources \citep[e.g.,][]{van-der-Burg:2016aa}.
Here, we highlight one of the largest overdensities of LSBGs, which can be seen
in bottom-left panel of Figure~\ref{fig:skypos} near $(\alpha, \delta)=
(225^\circ,\ +1^\circ)$. 

\begin{figure}[t!]
  \centering
  \includegraphics[width=\columnwidth]{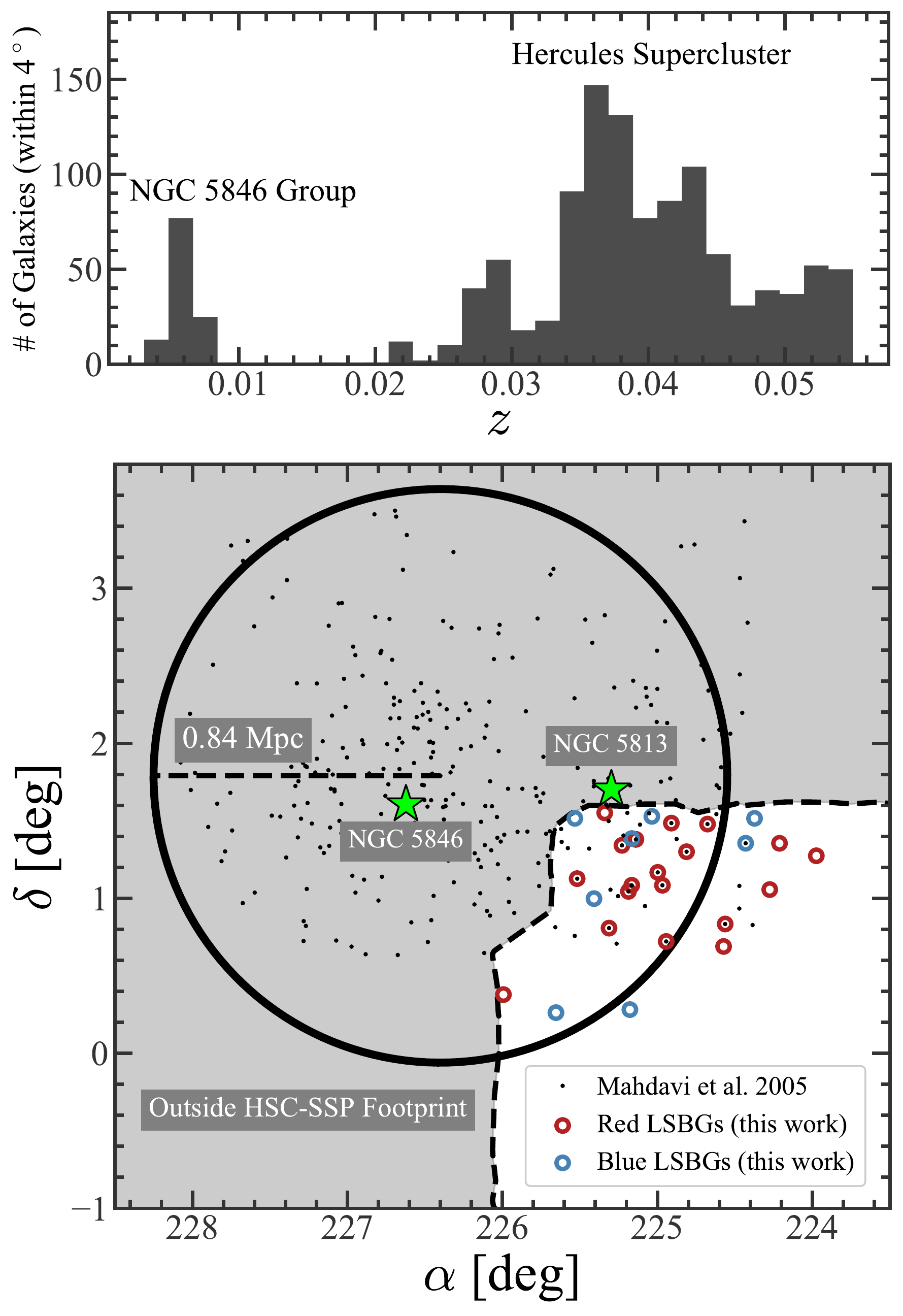}
  \caption{
    {\it Top:}
    Redshift distribution from the NSA catalog of all galaxies with known
    redshifts (with $z<0.055$) within $4^\circ$ (1.8~Mpc) of the NGC~5846
    group, which is remarkably isolated in space. The Hercules supercluster is
    behind this group and may host some of the LSBGs. 
    {\it Bottom:} 
    The sky positions of potential NGC~5846 group members (black points) with
    luminosities as faint as $M_R=-10$ \citep{Mahdavi:2005aa}.  The group is
    dominated by the giant ellipticals NGC~5846 and NGC~5813 (green stars). The
    white region in the bottom-right corner shows the \survey\ coverage near
    this group, and the open red (blue) circles show red (blue) LSBGs from this
    work. Sources also detected by \citet{Mahdavi:2005aa} have a black point
    within the open circle. The large black circle shows the boundary of the
    second turnaround radius at the distance of the group.  
  } 
  \label{fig:ngc-group} 
\end{figure}

Within this region, there are at least 27 LSBGs (19 red and 8 blue) clustered
within 1.5 times the projected second turnaround radius
\citep{Bertschinger:1985aa} of the NGC~5846 galaxy group, a nearby group at
26.1~Mpc \citep{Tonry:2001aa}. With a virial mass of
${\sim}8\times10^{13}~M_\odot$ \citep{Mahdavi:2005aa}, this group is the third
most massive collection of early-type galaxies in the local universe after the
Virgo and Fornax clusters, and it has been studied extensively at many
different wavelengths \citep[e.g.,][]{Tully:1987aa,Mulchaey:2003aa,
  Eigenthaler:2010aa,Machacek:2011aa,Marino:2016aa}. In the bottom panel of
Figure~\ref{fig:ngc-group}, we show the sky positions of potential group
members (black points) from a study of the galaxy population down to
luminosities as faint as $M_R=-10$ \citep{Mahdavi:2005aa}. Galaxies within this
group are predominantly of early type, and the system is dominated by the giant
ellipticals NGC~5846 and NGC~5813 (green stars). The large circle shows the
boundary of the second turnaround radius (0.84~Mpc) at the distance to the
group. The white region in the bottom-right corner of the figure shows the
coverage of the \survey, which only overlaps with a small fraction of the
group.

The open red (blue) circles in Figure~\ref{fig:ngc-group} show red (blue) LSBGs
from this work. Sources also detected by \citet{Mahdavi:2005aa} have a black
point within the open circle. As the figure shows, there is considerable
overlap between our catalogs. The \citet{Mahdavi:2005aa} sources within the
\survey\ footprint that are not in our catalog generally have much higher
surface brightness than our objects of interest; a couple were within the halo
of a bright star and were masked by our bright object mask.
\citet{Mahdavi:2005aa} detect 14 sources from our sample that overlapped with
their survey, most of which they assign as ``possible'' members\footnote{This
  corresponds to a score of 2 on a 0-4 system, with 0 being a spectroscopically
  confirmed member and 4 being likely not a member.  See
  \citet{Trentham:2001aa} for details about how membership probabilities were
  assigned.}. However, this is a statistical statement and does not apply to
any specific object, since none from our sample have been confirmed as members
via spectroscopic redshifts.  At the distance of the NGC~5846 group, our LSBGs
would have effective radii in the range $r_\mathrm{eff}\sim0.5$ to $1.6$~kpc
and absolute magnitudes in the range $M_g\sim-10.1$ to $-13.4$ mag, consistent
with dwarf elliptical galaxies. 

The NGC~5846 group is remarkably isolated in space. In the top panel of
Figure~\ref{fig:ngc-group}, we show the redshift distribution of galaxies from
the NSA catalog within $4^\circ$ (1.8~Mpc) of the group center. There is a
clear void between $z\sim0.01$-$0.02$, which is in stark contrast to the
background Hercules supercluster between $z\sim0.03$-$0.05$. Any LSBGs that lie
within this larger overdensity will have much more extreme physical properties.
We note that many of the LSBGs in Figure~\ref{fig:ngc-group} fall within the
projected virial radii of multiple cataloged galaxy groups at $z\sim0.04$ with
halo masses in the range $\log_{10}(M_\mathrm{halo}/M_\odot)\sim12.5$-$13.0$
\citep{Yang:2007aa}.

\subsection{A LSB supernova host}\label{sec:SN-host}

Novae and supernovae have been called ``beacons in the dark,'' which can be
used to detect LSBGs in the local universe \citep{Conroy:2015aa}. A galaxy in
our sample, LSBG-729, is a proof of concept for this idea. We show cutout
images of LSBG-729 from \survey, SDSS, and GALEX in the third column of
Figure~\ref{fig:galex}.  This galaxy hosted the Type IIb supernova SN 2009Z
\citep{Zinn:2012aa}, which was discovered by the Lick Observatory Supernova
Search \citep{Filippenko:2001aa} and classified by \citet{Stritzinger:2009aa}.
\citet{Zinn:2012aa} used SN 2009Z and LSBG-729 (called N271 in their work) to
demonstrate that it is possible to discover LSBGs via the observation of
ostensibly hostless supernovae. These authors used archival optical imaging
from SDSS and the 3.6~m New Technology Telescope to confirm the existence of
this LSB host. They then obtained an \HI\ spectrum of LSBG-729 using the 100 m
Effelsberg Radio Telescope, confirming that it is at a similar redshift as SN
2009Z ($z=0.02513$), and estimating an \HI\ mass of
$M_\mathrm{HI}=2.96\times10^9~M_\odot$. 

As can be seen in Figure~\ref{fig:galex}, LSBG-729 also has a strong
GALEX detection. Using Equation~(\ref{eqn:sfr}), we find a star-formation rate
of $\mathrm{SFR}= 0.09~M_\odot\ \mathrm{yr^{-1}}$. \citet{Zinn:2012aa} use SED
fitting to estimate that extinction from the galaxy itself may be as high as
$A_\mathrm{FUV}\approx2.5$. Accounting for this correction, we find a
relatively high star formation rate of $0.9~M_\odot$ yr$^{-1}$, comparable to
the Milky Way. Our estimates are about a factor of 2 larger than those of
\citet{Zinn:2012aa}, which may be attributed to deeper UV data (we use data
from GALEX's Medium Imaging Survey, whereas they use the All-Sky Imaging
Survey).

The \survey\ optical imaging is much deeper than was available to
\citet{Zinn:2012aa} (20 min exposures on an 8.2-m mirror compared to 1~min and
5~min exposures on 2.5- and 3.5-m mirrors, respectively). This is particularly
evident in Figure~\ref{fig:galex}, where the \survey\ image of LSBG-729 reveals
spiral structure and compact regions of star formation, whereas only its
central region is marginally visible in the SDSS image. Using
extinction-corrected \survey\ photometry and the same mass-to-light ratio/color
relation as above, we estimate a stellar mass of $M_\star\approx10^9~M_\odot$.
Using Equation~(\ref{eqn:f_gas}), this implies a gas fraction of
$f_\mathrm{gas}\approx0.8$. This fraction is lower than the estimate of
\citet{Zinn:2012aa}, which can be attributed to our larger stellar-mass
estimate. At $z=0.02513$, LSBG-729 has a large effective radius of
${\sim}4.6$~kpc.

\subsection{Size-Luminosity Relation}\label{sec:size-lum}

\begin{figure}[t!]
  \centering
  \includegraphics[width=\columnwidth]{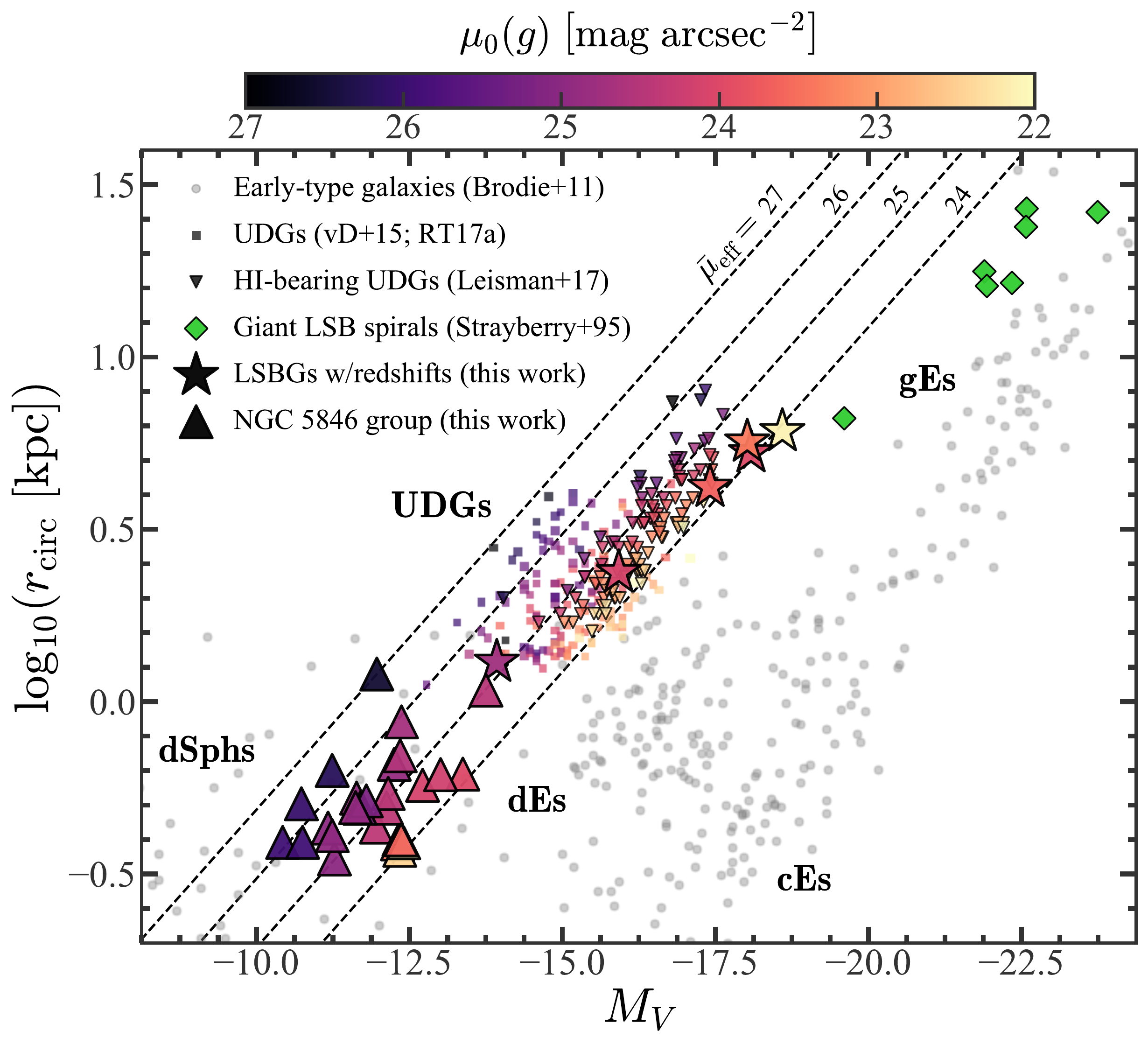}
  \caption{
    Size-luminosity relation for LSBGs in our sample for which we have distance
    information. Stars show LSBGs with archival spectroscopic redshifts, and
    large triangles show LSBGs that are projected in close proximity to the NGC
    5846 group (see Figure~\ref{fig:ngc-group}), which is at a distance of
    26.1~Mpc (we assume this distance for these LSBGs). We also show the family
    of early-type galaxies \citep{Brodie:2011aa}, giant LSB spiral galaxies
    \citep{Sprayberry:1995aa}, UDGs from \citealt{van-Dokkum:2015aa} (vD+15)
    and \citealt{Roman:2017aa} (RT167a), and \HI-bearing UDGs
    \citep{Leisman:2017aa}. The color bar shows the $g$-band central surface
    brightness for LSBGs in our sample and UDGs.  The y-axis shows the
    logarithm of the circularized effective radius $r_\mathrm{circ}
    =(1-\epsilon)^{1/2}\,r_\mathrm{eff}$. Lines of constant mean surface
    brightness are shown as dashed black lines.
  } 
  \label{fig:size-mag} 
\end{figure}

To summarize the distance information uncovered above: 6 LSBGs in our sample
have archival spectroscopic redshifts (LSBG-171, LSBG-456, LSBG-464, LSBG-575,
LSBG-613, and LSBG-729), and 27 LSBGs are projected in close proximity to a
nearby galaxy group at 26.1~Mpc, which we will assume as their distance.
Figure~\ref{fig:size-mag} shows the size-luminosity relation for these LSBGs.
\jpg{For context, we also show the family of early-type galaxies
  \citep{Brodie:2011aa}, representative giant LSB spiral galaxies
  \citep{Sprayberry:1995aa}, UDGs in intermediate to high density environments
  \citep{van-Dokkum:2015aa,Roman:2017aa}, and \HI-bearing UDGs in the field
  \citep{Leisman:2017aa}.} We convert $gr$ measurements to $V$-band using the
transformation $V=g-0.59\,(g-r)-0.01$ \citep{Jester:2005aa}. For the UDGs, we
assume $g-r=0.47$, which is the mean color observed by \citet{Roman:2017aa}. Galaxy
sizes are represented by the circularized effective radius $r_\mathrm{circ} =
(1-\epsilon)^{1/2}\,r_\mathrm{eff}$ (note the \citealt{Leisman:2017aa} objects
were measured within circular apertures due to low signal-to-noise ratio data).
 
Assuming that the 27 LSBGs (large triangles in Figure~\ref{fig:size-mag}) are 
at the same distance as the NGC 5846 group, they are physically similar to dwarf
spheroidals (dSphs) and dwarf ellipticals (dEs), with sizes ranging from
$r_\mathrm{circ} = 0.3$-1.2~kpc ($r_\mathrm{eff}=0.3$-1.6~kpc) and a median
absolute magnitude of $M_V=-12$. If some of these LSBGs are in fact associated
with the large overdensity behind this group (see Figure~\ref{fig:ngc-group}),
then their sizes and luminosities could potentially be much larger.

The 6 LSBGs with previous redshift measurements (stars in
Figure~\ref{fig:size-mag}) span a wide range in size-luminosity parameter
space---from small UDGs with $r_\mathrm{circ}=1.3$~kpc ($r_\mathrm{eff} =
1.6$~kpc) and $M_V=-14$ to giant LSB spirals with $r_\mathrm{circ}=6.1$~kpc
($r_\mathrm{eff}=6.7$~kpc) and $M_V=-19$. The largest/brightest of these
objects occupy the region of parameter space that falls between UDGs and giant
ellipticals, \jpg{similar to the lower luminosity end of previously known giant
  LSB spiral galaxies.} As indicated by the color bar in
Figure~\ref{fig:size-mag}, these large sources are among the higher surface
brightness objects in our sample (see Figure~\ref{fig:mu}). 

\section{Summary and Outlook}\label{sec:summary}

In this paper, we have presented our source-detection pipeline and an initial
catalog of sources from our ongoing search for extended low-surface-brightness
galaxies (LSBGs) in the Wide layer of the \survey. We have carried out our
search within the first ($gri$ full depth) \area~deg$^2$ of the survey, which
will extend to 1400~deg$^2$ upon completion.  Since our focus is on angularly
extended galaxies ($r_\mathrm{eff} = 2.5$-14\arcsec), our sample is likely
dominated by low-redshift sources.  We present
a catalog of \ngal\ LSBGs, where we define LSBGs in terms of mean surface
brightness $\left(\bar{\mu}_\mathrm{eff}(g)>24.3\ \mathrm{mag\
    arcsec^{-2}}\right)$, as opposed to central surface brightness $\mu_0$, to
allow nucleated galaxies into our sample. The contents of our 
catalog are summarized in Table~\ref{tab:catalog}.

We divide our LSBG sample into red ($g-i\geq0.64$) and blue ($g-i<0.64$)
galaxies, where the color boundary is at the median value. The surface
brightness distributions ($\bar{\mu}_\mathrm{eff}$ and $\mu_0$) are strong
functions of color, with the red distributions being much broader and generally
fainter than that of the blue LSBGs (Figure~\ref{fig:mu}); the median \mueff\
and $\mu_0(g)$ for red LSBGs are 25.8 and 24.9~\sbunit, respectively, whereas
they are 24.8 and 24.0~\sbunit\ for blue LSBGs.  Furthermore, this trend shows
a clear correlation with galaxy morphology (Figure~\ref{fig:stamps}). The
surface-brightness profiles of the red galaxies are typically very smooth and
well-characterized by single-component S\'{e}rsic functions (with median index
$n=0.9$). \jpg{Their morphologies and apparent clustering with galaxy groups
  (Figures~\ref{fig:skypos} and \ref{fig:ngc-group}) suggest they are composed
  of a combination of early-type dwarfs and UDGs in group
  environments---redshifts are required to distinguish between these
  possibilities for individual objects.} In contrast, the blue galaxies tend to
have irregular morphologies and show evidence of ongoing star formation.

We crossmatch our sample with archival data to gain insight into the physical
nature of the LSBGs (Section~\ref{sec:xmatch}). We find that our sample
encompasses a wide range of physical properties. From early-type dwarfs to
star-forming LSB spirals, our LSBGs span at least a factor of ${\sim}20$ and
${\sim}2000$ in physical effective radius and optical luminosity, respectively
(Figure~\ref{fig:size-mag}). Nearly half (46\%) of the galaxies in our sample 
(166 of which are blue) fall within the 
ALFALFA survey footprint; however, only 3 blue LSBGs have an \HI\ counterpart 
in the 70\% ALFALFA catalog (Section~\ref{sec:alfalfa}). 
Many of the blue LSBGs (296 out of 390), as well
some of the red LSBGs (78 out of 391), have UV detections in the GALEX source
catalog (Section~\ref{sec:galex} and Table~\ref{tab:catalog}). 

\resp{We note that we do not yet know the full redshift distribution 
(and thus the true surface-brightness distribution) of our 
sample. However, our minimum size cut at  $r_\mathrm{eff}=2\farcs5$ strongly 
biases our search to low-redshift sources. Consistent with this expectation, 
the redshifts we currently have in hand suggest a distance distribution with 
a range of ${\sim}30$-100~Mpc (Section~\ref{sec:xmatch}). 
In addition, a preliminary clustering analysis to be published in a future work 
yields a consistent, nearby distance distribution. Nevertheless, it is possible 
that a small fraction of objects in our sample may be 
normal- to high-surface-brightness sources at higher redshifts ($z\gtrsim0.1$), 
where cosmological dimming begins to become non-negligible. However, our 
sample pushes to very low central surface brightnesses, and as a result, 
even if our {\it entire} sample is at $z=0.1$ 
(which would imply all sources have $r_\mathrm{eff}>4.6$~kpc), ${>}93\%$ of 
our sources would have $\mu_0(g)>22$~\sbunit\ after correcting for 
cosmological dimming. It is, therefore, likely that the vast majority of 
galaxies in our catalog are true LSBGs.} 

The \survey\ is ushering in a new era for the study of LSBGs, which is
currently the best preparation we have for the even deeper and wider imaging
that will be produced by LSST. We consider this work a first step in the longer
term goal of building a statistical sample of ultra-LSB galaxies within the
complete \survey\ footprint. As such, we have not yet pushed the data as far as
they can go. For example, the sensitivity of our pipeline may be significantly
improved with a multi-scale approach such as wavelet decomposition
\citep[e.g.,][]{Prescott:2012aa}, or with new LSB-optimized
background-subtraction and/or shape-measurement algorithms. Both our current 
and future search methods may also be applied to the Deep and Ultra-Deep layers 
of HSC-SSP, which have limiting magnitudes that are ${\sim}$1 and 2 mag deeper 
than that of the Wide layer (over much smaller areas of sky). Nevertheless, 
our current LSBG catalog already demonstrates the potential of the \survey\ 
to deliver a truly unprecedented view of the galaxy population at low 
surface brightnesses.

Galaxies in the ultra-LSB regime represent a unique testing ground for
theoretical predictions of galaxy and star formation, stellar feedback
processes, and the distribution and nature of dark matter. The galaxy catalog
presented in this work will facilitate follow-up efforts to study the physical
properties and number densities of these elusive galaxies as a function of
environment. Pushing such studies to lower surface brightnesses will be
necessary to form a more complete census of the galaxy population, which will
ultimately provide one of the strongest tests of the standard $\Lambda$CDM
framework. 

\begin{table}[htb]
\caption{Low-surface-brightness galaxy catalog description} 
\label{tab:catalog}
\begin{tabular}{l|l|l}
\hline\hline
Column Name      & Unit    & Description                    \\
\hline
\multicolumn{3}{c}{Table: LSBGs (781 rows)}                 \\
\hline
id                       &         & Unique LSBG id \\
ra                       & deg     & Right ascension (J2000) \\
dec                      & deg     & Declination (J2000) \\
$\mu_0(i)$               & \sbunit & $i$-band central surface brightness \\
$\sigma(\mu_0(i))$       & \sbunit & Uncertainty of $\mu_0(i)$           \\
$m_i$                    & mag     & $i$-band apparent magnitude         \\
$\sigma(m_i)$            & mag     & Uncertainty of $m_i$            \\
$g-r$                    & mag     & $g-r$ color                     \\
$g-i$                    & mag     & $g-i$ color                     \\
$r_\mathrm{eff}$         & arcsec  & Effective radius                \\
$\sigma(r_\mathrm{eff})$ & arcsec  & Uncertainty of $r_\mathrm{eff}$ \\
$n$                      &         & S\'{e}rsic index                \\
$\sigma(n)$              &         & Uncertainty of $n$              \\
$\epsilon$               &         & Ellipticity                     \\
$\sigma(\epsilon)$       &         & Uncertainty of $\epsilon$       \\
$A_g$                    & mag     & $g$-band Galactic extinction \\
$A_r$                    & mag     & $r$-band Galactic extinction \\
$A_i$                    & mag     & $i$-band Galactic extinction \\
\hline
\multicolumn{3}{c}{Table: Archival GALEX Data (377 rows)}      \\
\hline
id                       &        & Unique LSBG id             \\
ra                       &  deg   & Right Ascension (J2000)    \\
dec                      &  deg   & Declination (J2000)        \\
NUV                      &  mag   & Near-UV apparent magnitude \\
$\sigma(\mathrm{NUV})$   &  mag   & Uncertainty of NUV         \\
FUV                      &  mag   & Far-UV apparent magnitude  \\
$\sigma(\mathrm{FUV})$   &  mag   & Uncertainty of FUV         \\
Survey                   &        & GALEX survey name          \\
\hline\hline
\end{tabular}
\tablecomments{
These tables are published in their entirety in machine-readable format.
Magnitudes are on the AB system and have not been corrected for Galactic
extinction. We provide Galactic extinction corrections, which are derived from
the \citet{Schlafly:2011aa} recalibration of the \citet{Schlegel:1998aa} dust
maps. The coordinates in the GALEX table are those of the matched source in
the GALEX source catalog (see Section~\ref{sec:galex}). 
}
\end{table}

\acknowledgments

We thank David Spergel and Jim Gunn for useful discussions about the galaxies
in our sample. J.P.G. thanks Jim Bosch and Paul Price for their assistance with
the LSST codebase, Adrian Price-Whelan for general coding advice, and Semyeong
Oh for useful conversations and for sharing data presented in
Figure~\ref{fig:ell}.  J.P.G. was supported by the National 
Science Foundation partially under grant No. AST 1713828 and partially 
through the Graduate Research Fellowship Program under Grant No. DGE 1148900. 

The Hyper Suprime-Cam (HSC) collaboration includes the astronomical communities
of Japan and Taiwan, and Princeton University. The HSC instrumentation and
software were developed by the National Astronomical Observatory of Japan
(NAOJ), the Kavli Institute for the Physics and Mathematics of the Universe
(Kavli IPMU), the University of Tokyo, the High Energy Accelerator Research
Organization (KEK), the Academia Sinica Institute for Astronomy and
Astrophysics in Taiwan (ASIAA), and Princeton University. Funding was
contributed by the FIRST program from Japanese Cabinet Office, the Ministry of
Education, Culture, Sports, Science and Technology (MEXT), the Japan Society
for the Promotion of Science (JSPS), Japan Science and Technology Agency (JST),
the Toray Science Foundation, NAOJ, Kavli IPMU, KEK, ASIAA, and Princeton
University. 

This paper makes use of software developed for the Large Synoptic Survey
Telescope. We thank the LSST Project for making their code available as free
software at http://dm.lsst.org.

The Pan-STARRS1 Surveys (PS1) have been made possible through contributions of
the Institute for Astronomy, the University of Hawaii, the Pan-STARRS Project
Office, the Max-Planck Society and its participating institutes, the Max Planck
Institute for Astronomy, Heidelberg and the Max Planck Institute for
Extraterrestrial Physics, Garching, The Johns Hopkins University, Durham
University, the University of Edinburgh, Queen‚Äôs University Belfast, the
Harvard-Smithsonian Center for Astrophysics, the Las Cumbres Observatory Global
Telescope Network Incorporated, the National Central University of Taiwan, the
Space Telescope Science Institute, the National Aeronautics and Space
Administration under Grant No. NNX08AR22G issued through the Planetary Science
Division of the NASA Science Mission Directorate, the National Science
Foundation under Grant No. AST-1238877, the University of Maryland, and Eotvos
Lorand University (ELTE) and the Los Alamos National Laboratory.

Based in part on data collected at the Subaru Telescope and retrieved from the
HSC data archive system, which is operated by Subaru Telescope and Astronomy
Data Center, National Astronomical Observatory of Japan.

Some of the data presented in this paper were obtained from the Mikulski
Archive for Space Telescopes (MAST). STScI is operated by the Association of
Universities for Research in Astronomy, Inc., under NASA contract NAS5-26555.
Support for MAST for non-HST data is provided by the NASA Office of Space
Science via grant NNX09AF08G and by other grants and contracts.

\software{
  Our source-detection pipeline is available at
  \url{https://github.com/johnnygreco/hugs} under an MIT open-source software
  license. This work additionally utilized \code{astropy}
  \citep{Astropy-Collaboration:2013aa}, \code{numpy}
  \citep{Van-der-Walt:2011aa}, \code{scipy} (\url{https://www.scipy.org}),
  \code{matplotlib} \citep{Hunter:2007aa}, \code{sqlalchemy}
  (\url{https://www.sqlalchemy.org}), \code{schwimmbad}
  \citep{schwimmbad}, \code{sfdmap} (\url{https://github.com/kbarbary/sfdmap}),
  and \code{photutils} \citep{photutils}.
}

\appendix 
\section{Low-Surface-Brightness False Positives} \label{sec:lsb-zoo}

The deep imaging afforded by \survey\ is essential for detecting new LSBGs;
however, it is also sensitive to other LSB phenomena such as Galactic cirrus
emission and tidal debris from galaxy interactions. These LSB sources, while
interesting in their own right, are common false positives in searches for
LSBGs. In this appendix, we provide examples of these LSB sources in data from
the \survey.

\subsection{Galactic Cirrus}

Optical scattered light from dust grains in the interstellar medium,
so-called Galactic cirrus, has long been recognized to exist
\citep{Elvey:1937aa,Guhathakurta:1989aa}, even at high Galactic latitudes
\citep{Sandage:1976aa}. Galactic cirrus can be a significant contaminant for
extragalactic studies, particularly when the focus is LSBGs or diffuse stellar
halos around massive galaxies \citep[e.g.,][]{Duc:2015aa}. At the same time,
the combination of optical imaging (which maps the cirrus on small
${\sim}1\arcsec$ scales) with infrared/microwave observations of
thermal emission (e.g., using WISE and/or {\it Planck}) is a powerful probe of
the physics of the interstellar medium
\citep[e.g.,][]{Miville-Deschenes:2016aa}.

\begin{figure}[htbp]
  \centering
  \includegraphics[width=\columnwidth]{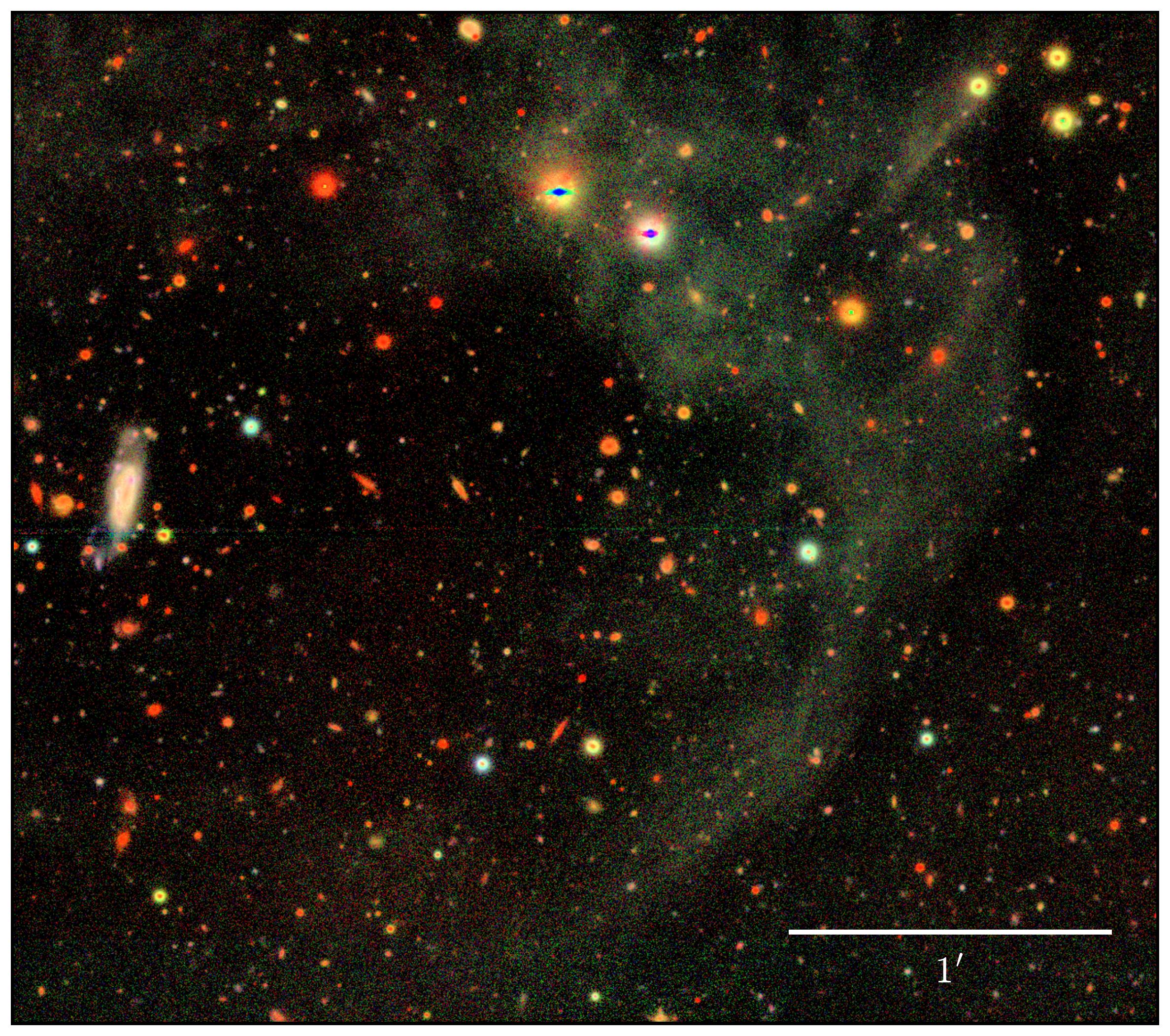}
  \caption{
    \survey\ $gri$-composite image of Galactic cirrus. Arising from the
    scattering of optical light off dust grains within the Milky Way, cirrus
    can be a source of contamination for extragalactic studies, even at high
    Galactic latitudes. The image is centered on $(\alpha,\
    \delta)=(332.5903^\circ,\ -0.516959^\circ)$, with north up and east to the
    left. 
  }
  \label{fig:cirrus}
\end{figure}

In general, Galactic cirrus is not a major source of contamination in our
search for LSBGs. One exception is within the \survey\ field that covers right
ascensions in the range $330^\circ<\alpha<345^\circ$ (bottom-right panel in
Figure~\ref{fig:skypos}). This field contains several patchy regions with large
amounts of optical cirrus, which have a filamentary character with structures
that span ${\sim}5$-$10^\prime$. In Figure~\ref{fig:cirrus}, we show a
$gri$-composite image of an example of Galactic cirrus within this field. Given the
characteristic wispy structure of Galactic cirrus and that cirrus clouds are
often found within large networks of similar clouds, it is generally
straightforward to eliminate such sources from our sample during our visual
inspection step.

\subsection{Tidal Debris}

At the depths of \survey, galaxy interactions produce rich networks of LSB
substructure and tidal debris, as predicted by the $\Lambda$CDM cosmological
framework \citep{Bullock:2005aa,Johnston:2008aa,Cooper:2010aa}. Therefore, the
\survey\ dataset offers the opportunity to study the build-up of massive
stellar halos across environments via large samples of faint tidal features
\citep[e.g.,][]{Atkinson:2013aa}. Such studies have the potential to constrain
the mass assembly rate of galaxies \citep{van-Dokkum:2005aa,Tal:2009aa} and
probe the orbital distributions of infalling satellite galaxies
\citep{Hendel:2015aa} throughout the universe. We will use \survey\ to 
study LSB tidal features in future work (Kado-Fong et al., in prep.).

Step~1 of our pipeline makes our search sensitive to LSB structures that are
relatively isolated from high-surface-brightness sources. As a result, a subset
of objects detected by our pipeline may be tidal debris from galaxy
interactions, which have been ejected far enough away from the primary system
to have a distinct footprint in our segmented detection images. We attempt to
identify and remove such sources during the visual inspection step of our
pipeline; however, there are inevitably ambiguous cases. For example, some
objects might be better classified as tidally disturbed satellites rather than
debris, and others may in fact be foreground dwarf galaxies with irregular
morphologies.  For a particularly interesting example of such an ambiguous
case, see \citet{Greco:2018aa}.

\begin{figure}[htbp]
  \centering
  \includegraphics[width=\columnwidth]{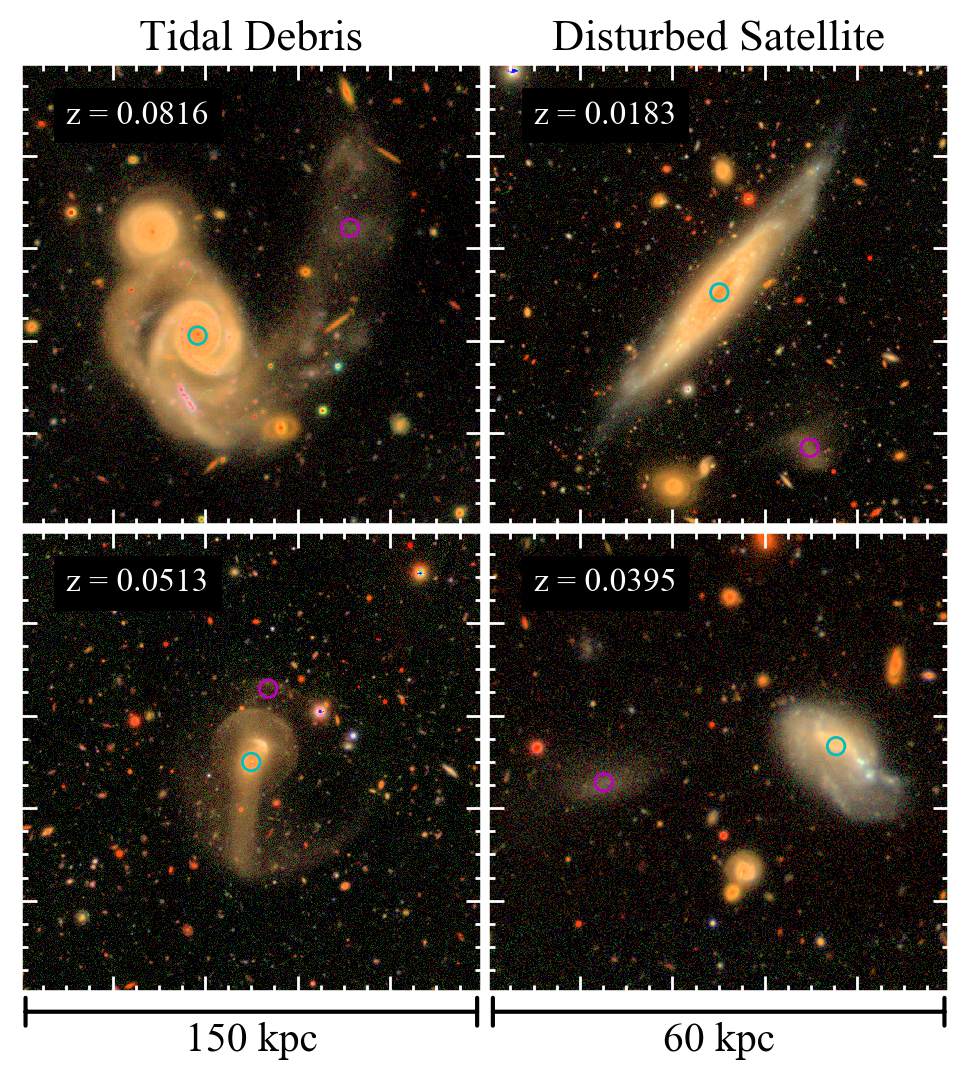}
  \caption{
    \survey\ $gri$-composite images of tidal debris (left column) and disturbed
    satellites (right column); note that the distinction between these types of LSB
    sources can be highly ambiguous. The sources detected by our pipeline are
    indicated by the magenta circles, and the likely host galaxies are
    indicated by the cyan circles. The physical scale for each column is shown 
    on the bottom, where each panel assumes the redshift of the host galaxy 
    indicated in the upper left. 
	}
	\label{fig:tidal}
\end{figure}

In Figure~\ref{fig:tidal}, we show $gri$-composite images of two objects we
classified as tidal debris (i.e., they are not in our final catalog; left
column) and two we classified as disturbed satellites (i.e., they are LSBGs in
our final sample; right column). We generally erred on the side of including 
possible tidal debris in the sample.

\bibliographystyle{apj}
\bibliography{ref}

\end{document}